\documentstyle[12pt,epsf]{article} 
\newcommand{\be}{\begin{equation}} 
\newcommand{\en}{\end{equation}}
\newcommand{\bea}{\begin{eqnarray}}
\newcommand{\ena}{\end{eqnarray}}

\newcommand{\hbo}{\hbox to 1 true cm {\hfill } } 
\newcommand{\tr}{\hbox{tr}}

\newcommand{\lan}{\Lambda _{I} } 
\newcommand{\laq}{\Lambda _{QCD} } 

\def\dslash{\partial\kern-.5em\slash}
\def\kslash{k\kern-.5em\slash}
\def\pslash{p\kern-.5em\slash}

\begin{document} 
\vglue 1truecm
  
\vbox{ UNITU-THEP-4/1996 
\hfill May 28, 1996
}
\vbox{ (revised) \hfill in press by Nucl. Phys. A}
  
\vfil
\centerline{\large\bf A renormalizable extension of the NJL-model$^*$ } 
  
\bigskip
\centerline{ K.\ Langfeld, C.\ Kettner$^1$, H.\ Reinhardt } 
\bigskip
\vspace{1 true cm} 
\centerline{ Institut f\"ur theoretische Physik, Universit\"at 
   T\"ubingen }
\centerline{D--72076 T\"ubingen, Germany.}
\bigskip
\vskip 1.5cm

\begin{abstract}
\noindent 
The Nambu-Jona-Lasinio model is supplemented by the quark interaction 
generated by the one-gluon exchange. The employed gluon propagator 
exhibits the 
correct large momentum behavior of QCD, whereas the Landau-pole at low 
energies is screened. The emerging constituent quark model is one-loop 
renormalizable and interpolates between the phenomenologically successful 
Nambu-Jona-Lasinio model (modified by a transversal projector) at low 
energies and perturbative QCD at high 
momenta. Consequently, the momentum dependence of the  quark self-energy 
at high energy coincides with the prediction from perturbative QCD. 
The chiral phase transition is studied in dependence on the low energy 
four quark interaction strength in the Dyson-Schwinger equation 
approach. The critical exponents  of the quark self-energy and the quark 
condensate  are obtained. The latter exponent deviates from the 
NJL-result. Pion properties are addressed by means of the Bethe-Salpeter 
equation. The validity of the Gell-Mann-Oakes-Renner relation is verified. 
Finally, we study the conditions under which the Nambu-Jona-Lasinio 
model is a decent approximation to our renormalizable theory as well as 
the shortcoming of the NJL-model due to its inherent non-renormalizability.

\end{abstract}

\vfil
\hrule width 5truecm
\vskip .2truecm
\begin{quote} 
$^*$ Supported in part by DFG under contract Re 856/1-3. 

$^1$ Supported by ``Graduiertenkolleg: Hadronen und Kerne''
\end{quote}
\eject
\section{ Introduction } 
\label{sec:1}

Most striking evidence for Quantumchromodynamics (QCD) as the right 
theory of strong interactions stems from high energy scattering 
experiments~\cite{yn83}, where experiments confirm the theoretical 
predictions. The remarkable property of asymptotic freedom implies 
that the running parameters of the theory, e.g.\ coupling constant and 
current masses, decrease with the inverse logarithm of the energy scale to 
the power of the corresponding anomalous dimension~\cite{yn83}, allowing 
a perturbative treatment. At low energies, the effective
coupling constant becomes large and even diverges at the so-called 
Landau pole which occurs at the typical energy scale of QCD 
$\laq$. It was recently argued~\cite{la95b}, that the 
Landau pole is an artifact of the assumption of a perturbative vacuum, 
and that non-perturbative physics, which sets in at low energies scales, 
screens the Landau singularity. Since non-perturbative QCD is not yet 
under control, this implies that for a description 
of low-energy hadron physics one has to resort to effective models. 

Low-energy hadron physics is dominated by the approximate chiral 
symmetry embodied by QCD~\cite{che84}. 
In their pioneering work, Nambu and Jona-Lasinio proposed 
an effective theory with chirally invariant four fermion 
interaction~\cite{nam61}, and 
illuminated the mechanism of spontaneous breaking of chiral symmetry, which 
led to a successful description of the light pseudo-scalar mesons as 
Goldstone bosons. The simple handling of the model allows for a wide range 
of applications describing the light mesons~\cite{vo91} and 
baryons~\cite{alk96}. Recently, the model was also extended to describe 
heavy hadrons~\cite{ro95}. 

Although the original version of Nambu and Jona-Lasinio 
was invented to describe nucleons and their interactions, the 
re-interpretation of the model with quarks as fundamental particles gained 
further importance, when it was realized that the low-energy effective 
quark theory of QCD should be of NJL type. This idea is supported in 
particular by the instanton picture~\cite{ca78} and the field strength 
approach to QCD~\cite{ha77,re91}. In the context of strong coupling QED, 
it was argued that the so-called ``irrelevant'' four fermion interaction 
of the effective theory becomes relevant due to a large anomalous 
dimension~\cite{bar86}. It was further observed that a NJL-type 
fermion interaction naturally arises in the dual formulation of strongly 
coupled scalar QED~\cite{che95}. These field theoretical investigations 
indicate that an effective quark model should 
exhibit a NJL-type interaction in the low energy regime. 

Despite the wide spread success of the NJL-model, it suffers from two 
shortcomings, i.e.\ the absence of quark confinement, which e.g.\ allows 
for a decay of the heavy mesons into free quarks, and its 
non-renormalizability. 
The latter enforces that the interaction is cut off at some momentum scale, 
which defines the range of applicability of the model. As a consequence, 
processes involving large momentum structure, 
e.g.\ form-factors at high momenta and anomalous decays like 
$\pi \rightarrow \gamma \gamma $, cannot be adequately described by the 
model. Physics which involve the interaction at high momentum can be properly 
described in the one-gluon exchange models~\cite{pa77,sme91,rob94}, 
which match the correct high energy behavior of QCD. 

The confining mechanism 
of QCD is still an unresolved problem. In certain super-symmetric Yang-Mills 
theories, it became apparent that the condensation of monopoles plays 
a key role~\cite{sei94}. Unfortunately, a concise understanding of 
confinement in QCD is not available at the moment.  At the level of an 
effective quark theory, an extension of the NJL-model was proposed which 
realizes confinement by random background fields~\cite{la95}. In the context 
of the one-gluon exchange models, confinement is described by a quark 
propagator which vanishes due to infra-red 
singularities~\cite{sme91}, or which does not have poles corresponding to 
asymptotic quark states~\cite{rob94}. In this paper, we will not address 
the question of confinement at the level of effective theories, 
but focus on the consequences of renormalizability\footnote{At the 
moment, there is little empirical evidence that confinement is 
important in hadron-physics at energy scales below the would-be 
quark thresholds.}. 

Here we propose an effective quark model which in the low energy regime 
coincides with the NJL-model modified by a transversal projector 
(see equation (\ref{eq:3})), and which matches the well known 
perturbative quark interaction at high energies at an energy scale 
$\lan$. We naturally choose $\lan > \laq $ in order that the 
NJL-part of the interaction screens the unphysical Landau pole. 
Since the interaction strength of the NJL-interaction is fixed by the 
constraint to match the interaction induced by the one-gluon exchange 
at high energies, the only parameter of our model is $\lan $. 

Our model should not be mixed up with the gauged 
NJL-model~\cite{bar86,co89,yam89}. In the latter, the NJL-interaction 
exists over the whole momentum range in addition to the gauge 
interactions. This model is non-perturbatively renormalizable and 
represents an interesting alternative to the standard mechanism of 
electro-weak symmetry breaking~\cite{mir89}. It is, however, not 
suitable in the context of the standard formulation of QCD. This is 
because its effective quark interaction does not reduce to the one-gluon 
exchange at high energies. 

The organization of the paper is as follows: In the next section, 
our model is defined, and its renormalization is discussed at the 
level of the Dyson-Schwinger equation. To this aim, the high-energy 
behavior of the quark self-energy is investigated by solving the 
Dyson-Schwinger equation numerically as well as analytically, 
using the Landau approximation in the latter case. 
Furthermore, the dependence of the chiral phase transition 
on the low energy four quark interaction strength is studied in some 
detail. In section \ref{sec:3}, pion physics is addressed by means 
of the Bethe-Salpeter equation. The Gell-Mann-Oakes-Renner relation 
is established. Numerical results concerning pion 
properties are presented. Finally, we compare our model with the standard 
NJL-model in section \ref{sec:4}. In particular, the shortcomings of the 
latter model due to its non-renormalizability are discussed. 
A summary and conclusions are 
left to the last section. Some explicit calculations are given in three 
appendices.

\goodbreak 
\section{ Model description and renormalization } 
\label{sec:2} 

\subsection{ Model building } 
\label{sec:2.1} 

For the description of hadrons we propose an effective one-gluon 
exchange model which is defined by the following generating functional 
(in Euclidean Space) 
\bea 
Z[\phi ] &=& \int {\cal D } q {\cal D} \bar{q} \; \exp \{ 
- \left[ S \, - \, \int d^4x \; \bar{q}(x) \phi (x) q(x) \right] \; , 
\label{eq:1} \\ 
S &=& \int d^4x \; \bar{q}( i \dslash + im ) q \, + \, \frac{1}{2}
\int d^4x d^4y \; \bar{q}(x) t^a \gamma _\mu q(x) 
{\cal D} _{\mu \nu }^{ab} (x-y) \bar{q}(y) t^b \gamma _\nu q(y) \; , 
\nonumber 
\ena 
where $m$ is the bare current quark mass, $t^a$, $a=1\ldots 8$ are the 
generators of SU(3) color group and $\gamma _\mu $ are the Hermitian 
Dirac matrices.  Furthermore, $\phi (x)$ is an external meson source. 
For $m=0$, the model is chiral invariant. 
The interaction is diagonal in color space, ${\cal D} _{\mu \nu }^{ab} = 
\delta ^{ab} D_{\mu \nu }$,  and consists of two parts, which, in 
momentum space, are separated by an energy scale $\lan $, i.e. 
\be 
D _{\mu \nu } (k) \; = \; 
\left[ G \, \theta (\lan - \vert k \vert ) \; + \; 
\frac{4 \pi \alpha (k^2) }{ k^2 } \theta (\vert k \vert - \lan ) \right] 
\, \left( \delta _{\mu \nu } - \frac{ k_\mu k_\nu }{k^2} \right) \; , 
\label{eq:3} 
\en 
where $\theta (x)$ is the step function and $\vert \cdot \vert $ denotes 
the module at a Euclidean 4-vector. Choosing 
\be 
\alpha (k^2) \; = \; \frac{12 \pi }{ (33-2 N_f) \ln k^2/\laq ^2} \; , 
\label{eq:4} 
\en 
where $N_f$ is the number of quark flavors, the second term in 
(\ref{eq:3}) represents for $k>\lan $ the perturbative gluon propagator 
of QCD multiplied by the effective (running) coupling constant
$\alpha (k^2)$. The coupling strength $G$ of the 
NJL-term is chosen to match the gluonic interaction at the scale $\lan $, 
i.e. 
\be 
G \; = \; \frac{ 4 \pi \, \alpha (\lan ^2) }{ \lan ^2 } \; . 
\label{eq:4a} 
\en 

In order to study the spontaneous breakdown of chiral symmetry, we 
investigate the quark self-energy $\Sigma (p)$. It can be obtained 
from the Dyson-Schwinger equation, which in ladder approximation reads
\be 
-i \pslash \, + \, m \, + \, \frac{4}{3} 
\int \frac{ d^4k }{(2\pi )^4 } \; \gamma _{\mu } \, S(k+p) \, 
\gamma _\nu \; D_{\mu \nu }(k) \; = \; S^{-1}(p) \; . 
\label{eq:5} 
\en 
Here 
\be 
S(p) \; = \; \frac{ i }{ Z(p^2) \pslash \, + \, i \Sigma (p) } \; . 
\label{eq:6} 
\en 
is the quark propagator, which is diagonal in color space, and we have 
used \break $\sum_{a,l} t^a_{il} t^a_{lk} = \frac{4}{3} \, 
\delta _{ik}. $ 
It will turn out (see section \ref{sec:2.3}) that the loop integration 
in (\ref{eq:5}) is in fact ultra-violet finite, since the gluon 
propagator decreases like $1/k^2 \, \ln (k^2 / \laq ^2)$ for large 
values of $k^2$, and the self-consistent solution $\Sigma (k^2) $ 
is asymptotically proportional to $1/[\ln k^2/\mu^2 ]^{d_m}$, where 
$d_m = 4/9$ is the anomalous mass dimension for three quark flavors. 
In addition, no finite renormalization constants 
are needed in (\ref{eq:5}), because the freedom in the functions $Z(p^2)$ 
and $\Sigma (p^2)$ is sufficient to accomplish for the standard 
renormalization conditions of the quark propagator. 

Using the ansatz (\ref{eq:6}), the Dyson-Schwinger equation can be cast 
into two equations which determine the self-energy $\Sigma (p^2)$ and 
the quark wave-function $Z(p^2)$, i.e. 
\bea 
\Sigma (p^2) &=& m \, + \, 4G \int _{\vert k \vert < \lan } 
\frac{ d^4k }{(2\pi )^4 } \; \frac{ \Sigma ((p+k)^2) }{ 
Z^2((p+k)^2) (p+k)^2 + \Sigma ^2((p+k)^2) } 
\label{eq:7} \\ 
&+& 4 \int _{\vert k \vert > \lan } 
\frac{ d^4k }{(2\pi )^4 } \; \frac{ \Sigma ((p+k)^2) }{ 
Z^2((p+k)^2) (p+k)^2 + \Sigma ^2((p+k)^2) } \; 
\frac{ 4 \pi \alpha (k^2) }{ k^2 } \; , 
\nonumber \\ 
Z(p^2) &=& 1 + \frac{4G}{3} \int _{\vert k \vert < \lan } 
\frac{ d^4k }{(2\pi )^4 } \, \frac{ Z ((p+k)^2) }{ 
Z^2((p+k)^2) (p+k)^2 + \Sigma ^2 } \left( 1 + 3 \frac{p \cdot k }{p^2} 
+2 \frac{ (p \cdot k)^2 }{k^2 p^2} \right) 
\nonumber \\ 
&+& \frac{4}{3} \int _{\vert k \vert > \lan } 
\frac{ d^4k }{(2\pi )^4 } \; \frac{ Z ((p+k)^2) }{ 
Z^2((p+k)^2) (p+k)^2 + \Sigma ^2 } \left( 1 + 3 \frac{p \cdot k }{p^2} 
+2 \frac{ (p \cdot k)^2 }{k^2 p^2} \right) \, 
\nonumber \\ 
&\times & \frac{ 4 \pi \alpha (k^2) }{ k^2 } \; . 
\label{eq:8} 
\ena 
Note, that the $k$-integration never crosses the Landau singularity 
at $\vert k \vert = \laq $ due to our assumption $\lan > \laq $.

\subsection{ Numerical results } 
\label{sec:2.2} 

To pave the ground for later investigations, we first 
present a numerical solution of the Dyson-Schwinger equation 
in the chiral limit $m \equiv 0$ and discuss its relevant properties. 
The numerical method to solve the system of coupled differential 
equations as well as the procedure to estimate the numerical 
errors are described in some detail in appendix \ref{app:a0}. 

The actual choice of the fundamental energy scale $\laq $ 
depends on the regularization scheme, which is used to extract the 
divergent parts. In the $\overline{MS}$-scheme, which employs dimensional 
regularization, its typical value is $\laq \approx 100 \ldots 
200 \, $MeV~\cite{yn83}, whereas in the one-gluon exchange models, 
which generically use a sharp O(4)-invariant cutoff (and therefore a 
momentum-subtraction scheme), values up to 
$500 \, $MeV are consistent with phenomenology~\cite{pa77,sme91}. 
In the case of the present model, it will turn out that the choice 
$\laq = 500 \, $MeV reproduces the experimental data for the 
static pion properties (see section \ref{sec:3}). 
For illustrative purposes, let us choose $G = 1.11 \; 16\pi^2/ \laq ^2 $. 
It will turn out, that this choice for $G$ is to small to provide 
a constituent quark mass of $M \approx 310 \, $MeV. 

\begin{figure}[t]
\centerline{ 
\epsfxsize=9cm
\epsffile{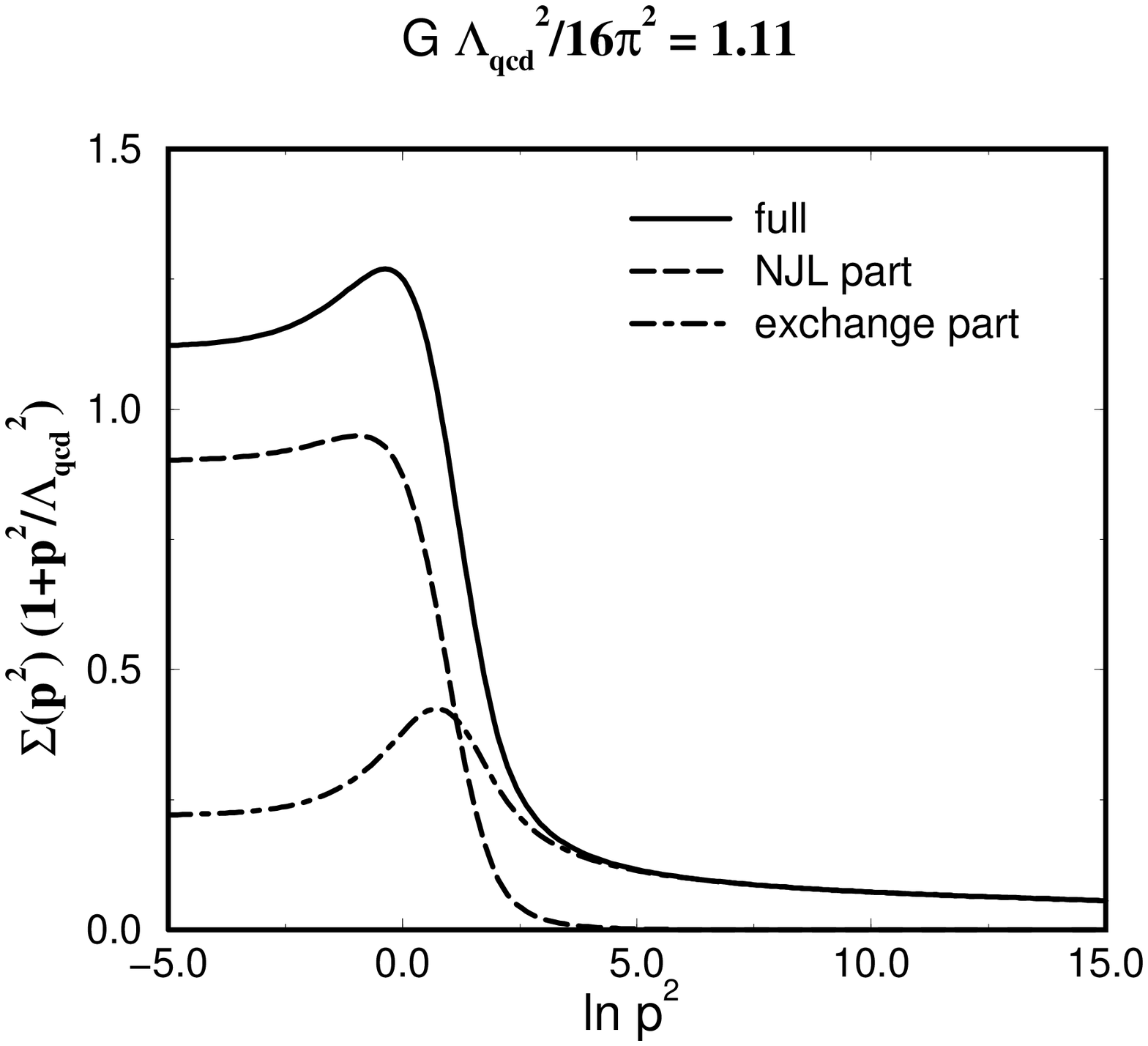} 
\epsfxsize=9cm
\epsffile{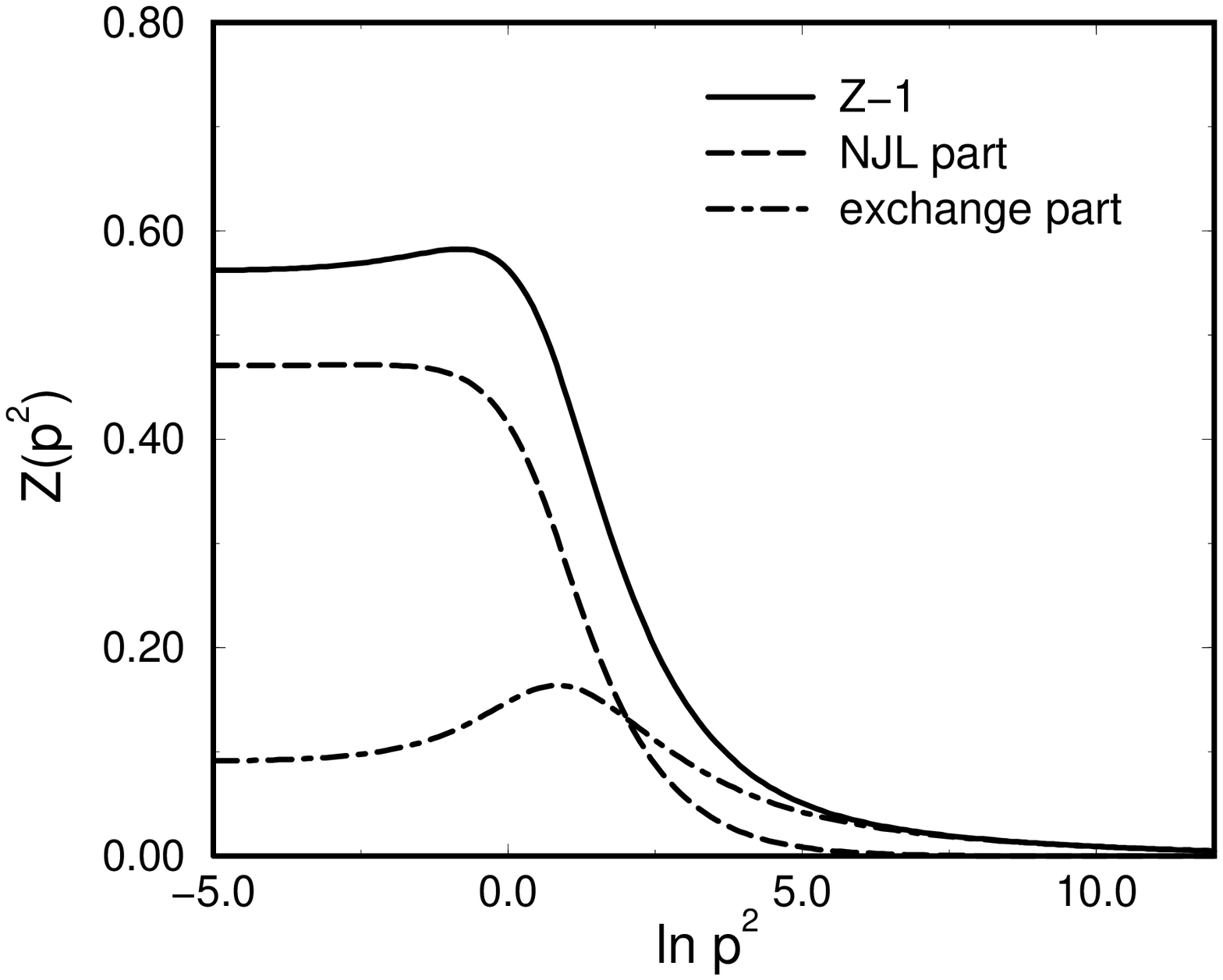} 
}
\vspace{.5cm} 
\caption{ The self-energy $\Sigma (p^2)$ times $1+p^2/\laq^2$ in units 
   of $\laq$ 
   and the wave function $Z(p^2)$ as function of the momentum transfer 
   $p^2$ (full lines). Also shown are the contributions from the 
   NJL part of the interaction (long dashed) and from the gluon exchange 
   term (dot-dashed). } 
\label{fig:1} 
\end{figure}

The numerical result for the self-energy $\Sigma (p^2)$ and the 
wave function $Z(p^2)$ for three flavors ($N_f=3)$ is shown in 
figure~\ref{fig:1}. 
Although both integrals in (\ref{eq:7}) are equally important in order 
to calculate the self-consistent solution $\Sigma (p^2)$, it is 
instructive to split $\Sigma (p^2)$ into the contributions stemming from 
the NJL type interaction (first line in (\ref{eq:7})) and from the 
one-gluon exchange term {\it after} the self-consistent solution was 
obtained. These contributions are also plotted in figure \ref{fig:1}. 
As expected, the NJL term dominates at small momentum transfer, 
whereas the high momentum behavior is exclusively determined by the 
one-gluon exchange part of the interaction.

\begin{figure}[t]
\centerline{ 
\epsfxsize=12cm
\epsffile{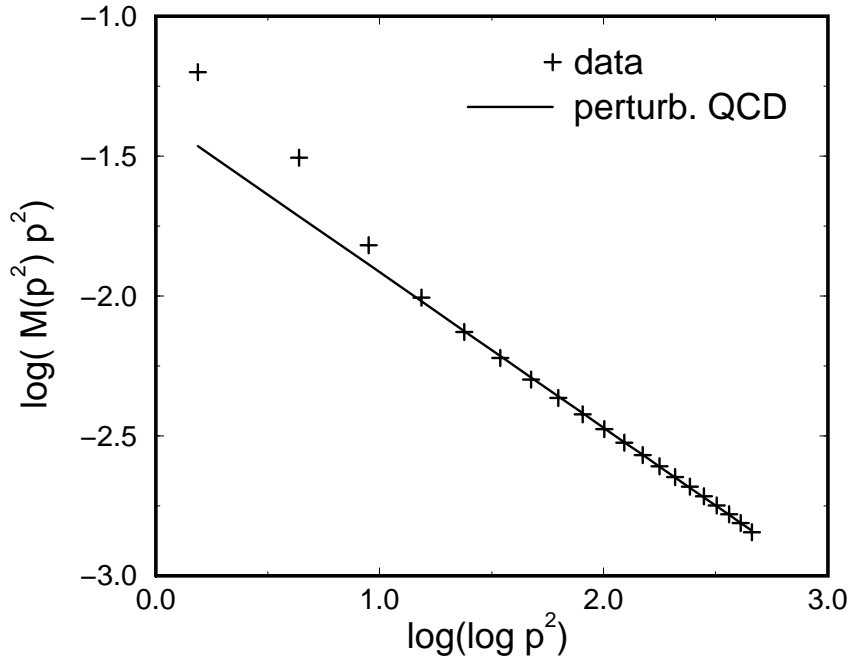} 
}
\vspace{.5cm} 
\caption{ The asymptotic behavior of the quark mass function $M (p^2) $ 
   (points) compared with the result of perturbative QCD (line). }
\label{fig:2} 
\end{figure} 
The high-momentum behavior of the quark constituent mass function, i.e. 
\be 
M(p^2) \; := \; \frac{ \Sigma (p^2) }{ Z(p^2) } \; , 
\label{eq:9a} 
\en 
is of particular interest, since it can be compared with the result 
from perturbative QCD. Including the leading  operator-product 
corrections, the mass function is asymptotically (for large values of $p^2$) 
given by~\cite{pol76}, 
\be 
M (p^2) \approx \frac{ m_R (\mu ) \left[ 
\ln ( \mu ^2 / \laq^2 ) \right] ^{d_m} }{ \left[ 
\ln ( p ^2 / \laq^2 ) \right] ^{d_m} } \, - \, 
\frac{ 4 \pi ^2 d_m }{3p^2} \frac{ \langle \bar{q} q \rangle 
(\mu ) \left[ \ln ( \mu ^2 / \laq^2 ) \right] ^{-d_m} 
}{ \left[ \ln ( p ^2 / \laq^2 ) 
\right] ^{1-d_m} } \; + \; \cdots 
\label{eq:10} 
\en 
where $m_R$ and $ \langle \bar{q}q \rangle $ is the renormalized 
current mass and the quark condensate, respectively, at a 
renormalization point $\mu $, and $d_m$ is the anomalous dimension of the 
current mass. In perturbative QCD one finds~\cite{yn83,pol76} 
\be 
d_m \; = \; \frac{12}{ 33 - 2 N_f } \; = \; \frac{4}{9} \; 
\hbox to 2 cm {\hfil for \hfil } N_f=3 \; . 
\label{eq:11} 
\en 
In the chiral limit, the mass function $M(p^2)$ asymptotically 
decreases like $1/p^2$ with logarithmic corrections. The same is true 
for the self-energy $\Sigma (p^2)$, since the 
wave function $Z(p^2)$ logarithmically approaches $1$. 
Figure \ref{fig:2} exhibits the asymptotic behavior of the numerical 
result for $M (p^2)$. 
The numerical data show a logarithmic decrease of $p^2 \, 
M (p^2)$. The line corresponds to an anomalous dimension 
of the quark condensate of $ 1- d_m = 5/9 $, which agrees with the result 
of perturbative QCD~\cite{yn83}.

\subsection{ Analytical results } 
\label{sec:2.3} 

The asymptotic behavior of the quark mass function (\ref{eq:10}) 
is well known from perturbative QCD studies augmented with 
OPE corrections~\cite{pol76}. In the following we will show that 
our model Dyson-Schwinger equation (\ref{eq:7},\ref{eq:8}) yields a
quark self-energy $\Sigma (p^2)$ which has the correct asymptotic 
(large $p^2$) behavior predicted by QCD. The asymptotic form 
of Dyson-Schwinger equations was studied in detail for many 
models. For a review we refer to the book of Miransky~\cite{mir93}. 

In the case of our model, we firstly show that the NJL part of the 
integral equation (\ref{eq:7}) does not affect the asymptotic behavior 
of $\Sigma (p^2)$, as it was already observed in the numerical 
calculation (see figure \ref{fig:1}). In fact, since the momentum 
integration in the NJL-term is over a finite volume in momentum space, 
the integrand can be expanded in powers of $k^2/\lan ^2 $. 
Obviously, this contribution to $\Sigma (p^2)$ is suppressed by powers of 
$\lan ^2/ p^2$, i.e. 
\be 
\Sigma (p^2) \; = \; 
\frac{ G }{ 8 \pi ^2 } \; \frac{ \lan ^4 }{ Z^2(p^2) p^2 } 
\; \Sigma (p^2) \; + \; \ldots \; , 
\label{eq:13} 
\en 
implying that asymptotically ($p^2 \rightarrow \infty$) only the 
gluon exchange term contributes. 

In appendix \ref{app:aa} we show that in the Landau 
approximation\footnote{ This approximation is only applied in this 
subsection. The numerical studies in this paper treat the full 
Dyson-Schwinger equations (\ref{eq:7},\ref{eq:8})} and for 
non-vanishing renormalized current mass the coupled set of 
Dyson-Schwinger equations (\ref{eq:7},\ref{eq:8})  has for $p^2 
\rightarrow \infty $ the asymptotic solution 
\be 
\Sigma (p^2) \; \approx \; \frac{ \kappa }{ \left[ \ln \frac{ p^2 }{ \laq ^2} 
\right] ^{\alpha } } \; , 
\label{eq:12} 
\en 
where 
\be 
\alpha \; = \; \frac{12}{33 - 2N_f} \; = \; d_m \; , 
\en 
and $\kappa $ is proportional to the renormalized current mass $m_R$. 
Furthermore, the bare current mass runs with the cutoff as 
\be 
m(\Lambda _{UV}) \; = \; \frac{ \kappa }{ [ \ln \Lambda _{UV}^2 
/ \laq ^2 ]^\alpha } \; , \hbo 
\kappa \; = \; 
m_R (\mu ) \left[ \ln ( \mu ^2 / \laq^2 ) \right] ^{d_m} \; . 
\label{eq:21} 
\en 
In the chiral limit $(m_R=0)$, the asymptotic solution is given by 
\be 
\Sigma (p^2) \; \approx \; \frac{ c }{ p^2 \left[ \ln \frac{ p^2 }{ \laq ^2} 
\right] ^{\beta } } \; , \hbo \beta \, = \, 1 - d_m \; , 
\label{eq:12a} 
\en 
where $c$ is measure for the quark condensate (compare with equation 
(\ref{eq:10})). 

The main observation in this subsection is that the large momentum 
behavior of the solutions $\Sigma (p^2)$, $Z(p^2)$ of the coupled 
set (\ref{eq:7},\ref{eq:8}) of differential equations agrees with 
the QCD predictions.

\subsection{ Chiral phase transition } 
\label{sec:2.4} 

\begin{figure}[t]
\centerline{ 
\epsfxsize=9cm
\epsffile{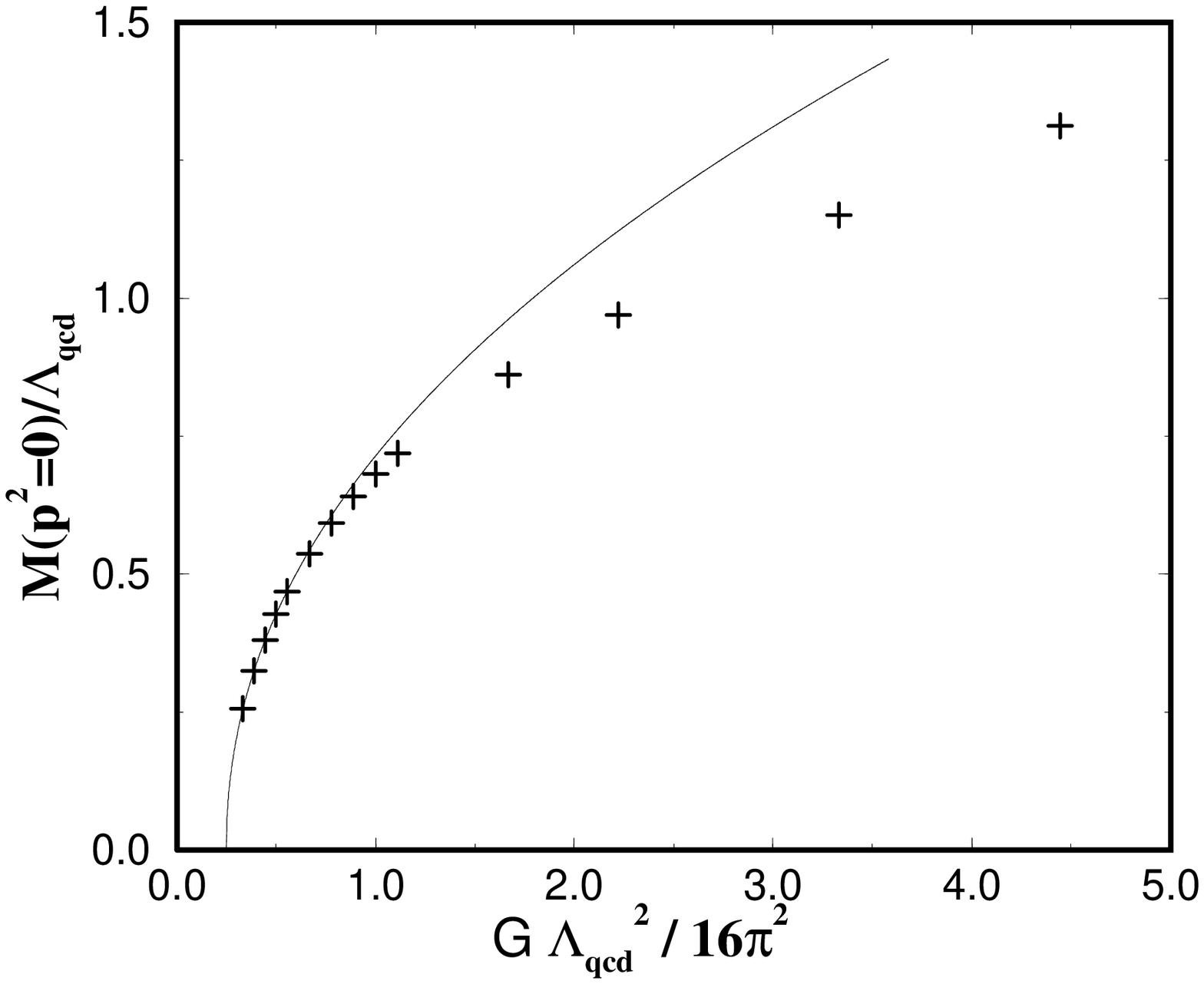} 
\epsfxsize=9cm
\epsffile{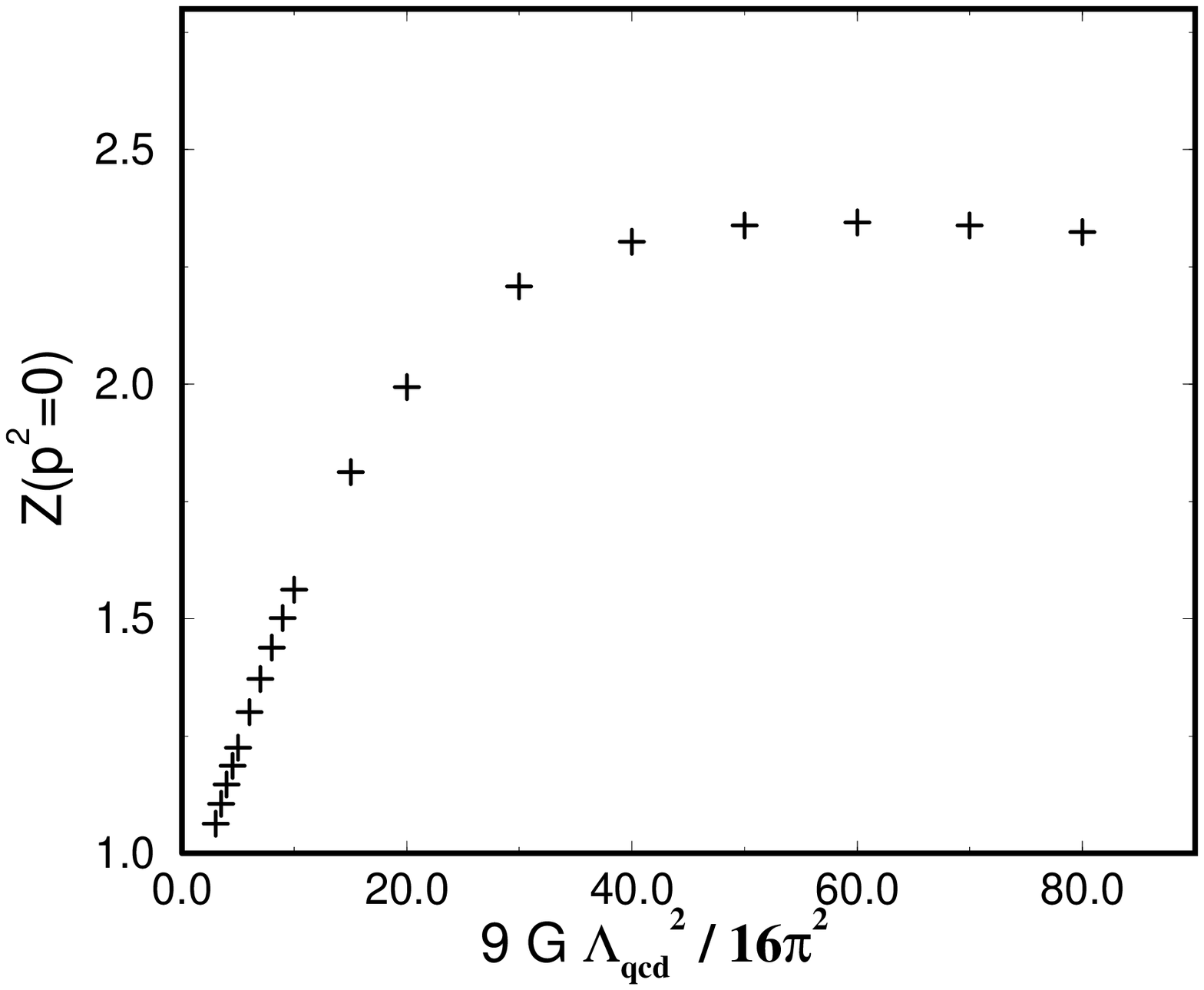} 
}
\vspace{.5cm} 
\caption{ The quark mass function $M (p^2=0)$ (crosses) and $Z(p^2=0)$ 
   (crosses) at zero momentum transfer as function of the coupling 
   strength $G$ of the low energy effective four quark interaction. 
   Also shown is the fit (solid line) according (\protect{\ref{eq:26}}).} 
\label{fig:3} 
\end{figure} 
The results found in the previous subsection, i.e. a non-vanishing quark 
self-energy with the correct decrease at asymptotic large values of the 
momentum transfer, establishes that chiral symmetry is spontaneously broken 
for an appropriate choice of parameters $(G>G_c)$. Here, we will report 
details of the chiral phase transition at the critical value $G_c$ of the 
effective four quark interaction. We have numerically 
investigated the quark self-energy $\Sigma (p^2=0)$ and the wave function 
$Z(p^2=0)$ at zero momentum transfer by solving the coupled 
set of integral equations (\ref{eq:7}) and (\ref{eq:8}) for various 
values of $G$, 
the only parameter\footnote{ Note that $\lan $ is fixed by the constraint 
(\ref{eq:4a}).} which is not fixed by perturbative QCD results. 
The results are depicted in figure \ref{fig:3}. For small values 
of $G<G_c$, the chiral symmetry is not spontaneously broken 
and $\Sigma (p^2)=0$ is the only solution to equations 
(\ref{eq:7},\ref{eq:8}). However, if the coupling exceeds the critical 
value\footnote{ The procedure to obtain an estimate of the numerical error 
is discussed in appendix \protect{\ref{app:a0}}.} 
\be 
G_c \approx \biggl( 0.26 \, \pm \, 0.011 \biggr) \; \times \; 
\frac{16 \pi ^2 }{ \laq ^2 } \; , 
\label{eq:25} 
\en 
a non-trivial self-energy signals the spontaneous breakdown of chiral 
symmetry. For values $G$ slightly above the critical coupling $G_c$, 
our numerical results are well fitted by 
\be 
M (p^2=0) \; \approx \; 0.43 \, \laq \; \left( \frac{G}{G_c}-1 
\right) ^{\delta _\Sigma } \;  \hbo 
\delta _{\Sigma } \, = \, 0.47 
\; . 
\label{eq:26} 
\en 
If $\epsilon = G/G_c-1$ measures the deviation of the coupling strength 
from its critical value, the behavior of the order parameter 
close to the phase transition is in general $M \approx 
\epsilon ^\delta $, where the critical exponent is 
characteristic of the underlying field theory only~\cite{zinn}. 
In our case (\ref{eq:26}), a critical exponent of $1/2$ is compatible 
with the numerical data. 

Figure \ref{fig:3} also shows the wave function $Z(p^2=0)$ as function 
of $G$. For $G<G_c$, the wave function takes its 
canonical value $Z(p^2=0)=1$. For values $G$ larger than the critical 
coupling $G_c$, the function $Z(p^2=0)$ linearly increases with 
increasing strength $G$. It saturates, and becomes a constant independent 
of $G$ for sufficiently large $G$. 

\begin{figure}[t]
\centerline{ 
\epsfxsize=9cm
\epsffile{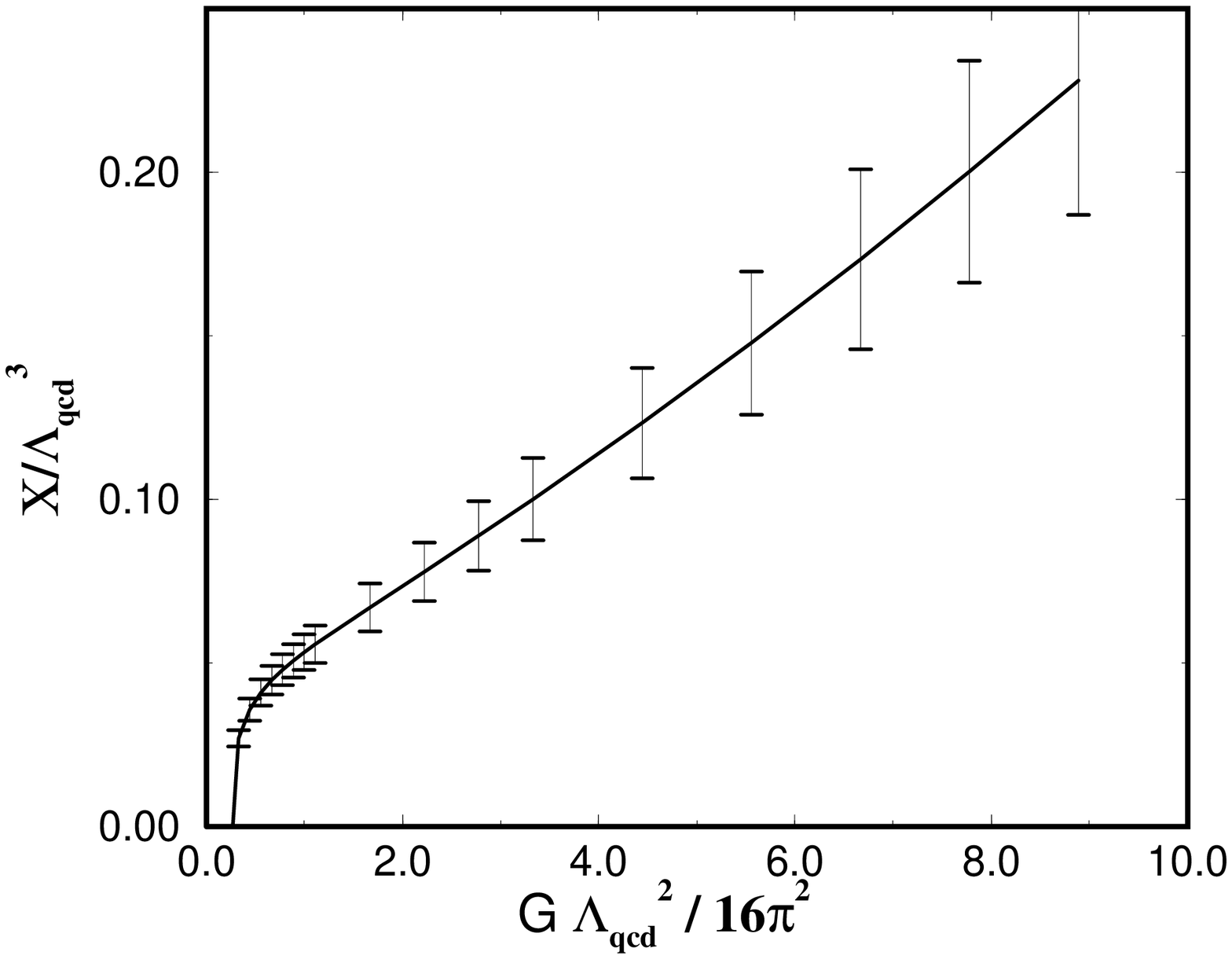} 
\epsfxsize=9cm
\epsffile{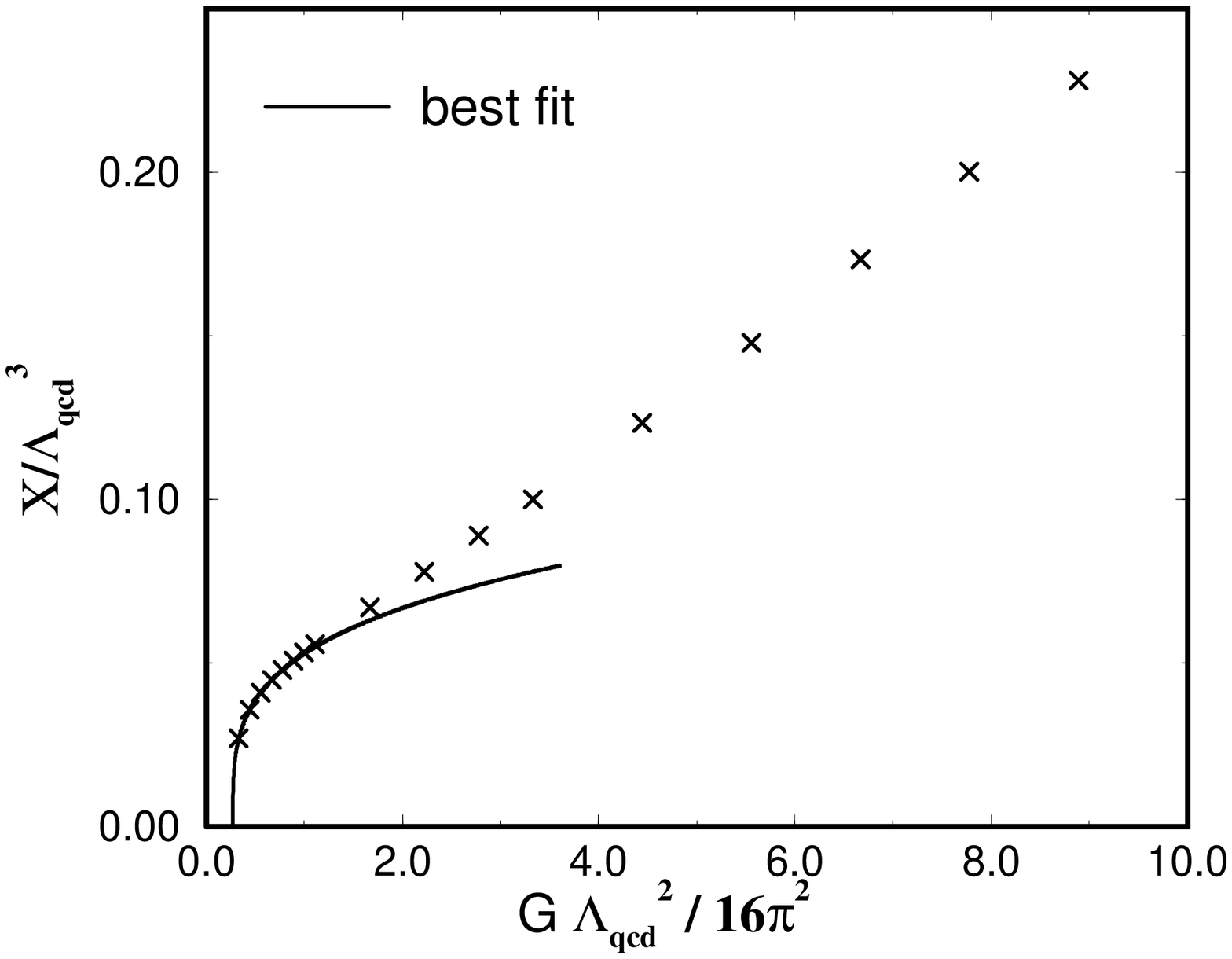} 
}
\vspace{.5cm} 
\caption{ The quark condensate $\chi = - \langle \bar{q} q \rangle $ 
   at a renormalization point 
   $\mu = 1\, $GeV as function of the low energy interaction strength $G$. 
   The solid line in the right hand picture is the fit function 
   (\protect{\ref{eq:28}}). } 
\label{fig:4} 
\end{figure} 
Finally, we present the quark condensate as a function of the interaction 
strength $G$. The quark condensate at a renormalization point $\mu $ 
is defined via the trace of the quark propagator $S(k)$ (\ref{eq:6})
(see subsection \ref{sec:3.2}), i.e. 
\be 
\langle \bar{q} q \rangle ( \mu ) \; = \; 
- \, \left[ \ln \frac{ \mu ^2 }{ \laq ^2 } \right] ^{d_m} \; 
\left[ \ln \frac{ \Lambda _{UV}^2 }{ \laq ^2 } \right] ^{-d_m} \; 
\int _{k^2 < \Lambda _{UV}} \frac{ d^4k }{(2 \pi )^4} \; \tr \, S(k) \; . 
\label{eq:27} 
\en 
It was shown in~\cite{la96} that this definition of the quark condensate 
is compatible with the definition via the asymptotic behavior of 
the mass function (\ref{eq:10}), which is used in the operator 
product expansion (see also subsection \ref{sec:3.2}). 
The numerical result of the quark condensate at 
a renormalization point of $\mu = 1 \, $GeV is shown in figure \ref{fig:4}. 
For values $G$ close to $G_c$, the numerical data is well represented by 
the function 
\be 
\langle \bar{q} q \rangle (\mu = 1\, \hbox{GeV}) \; \approx \; 
- 0.041 \; \laq ^3 \; \left( \frac{G}{G_c} -1 \right) ^{\delta _{\chi } }
\; , \hbox to 2 cm {\hfil with \hfil } \delta _\chi \approx 0.27 \; . 
\label{eq:28} 
\en 
In the standard NJL-model, the quark 
condensate is directly proportional to the self-energy implying that 
this model yields a critical exponent of $1/2$. This simple connection 
between the self-energy and the condensate is artificial, since the 
condensate is in general a subject of renormalization. In particular, 
the critical exponent of our model significantly deviates from $1/2$. 

\section{ Static pion properties } 
\label{sec:3} 

\subsection{ The Bethe-Salpeter equation } 
\label{sec:3.1} 

In the constituent quark picture, the pion is described as 
a pseudo-scalar bound state of a quark anti-quark pair. This bound state 
is characterized by its vertex function $P(p,q)$, which satisfies a 
Bethe-Salpeter 
(BS) equation. The momentum $p$ thereby represents the four 
momentum of the pion, and $q$ reflects the internal structure of the 
pion. In our case, the BS-equation is 
\be
P(p,q) \ = \ \int_k i \gamma_\mu \ t^a \ D_{\mu \nu} (k-q) \  S(k+p/2) \
P(p,k) \ S(k-p/2) \ t^a \ i \gamma_\nu \; , 
\label{eq:30} 
\en 
where $D_{\mu \nu }(k-p)$ is the effective interaction (\ref{eq:3}) of 
our model. 

In the following, we will solve this equation (\ref{eq:30}) within a 
derivative expansion. This expansion was earlier obtained in 
references~\cite{cah96} and~\cite{fra96}. 
For later use, we briefly re-derive the expansion in our notation. 

For this purpose, first note that the BS-equation (\ref{eq:30}) 
can be formally written as an eigenvalue equation, i.e. 
\be
P_p  \  = \ \int_q \ K(\Sigma , p^2, k, q)  \ P_p(q)
\quad  \mbox{ or : }  \quad P_p =   K(\Sigma , p ) P_p \ , 
\label{eq:31} 
\en 
where the BS-vertex $P_p$ is represented as a vector, and the 
kernel $K(\Sigma , p )$ depends on the self-energy $\Sigma (k^2)$ and 
the pion momentum $p$. For later convenience, we introduce a 
``scalar product'' of two vectors $a$ and $b$ via a quark loop 
integral, i.e. 
\be
\langle  a \, \vert \, b \rangle := \tr \ \int 
\frac{ d^4k }{(2\pi )^4} \; S_0(k-\frac{p}{2}) \ a^\dagger 
(k-\frac{p}{2},k+\frac{p}{2}) \ 
S_0(k+\frac{p}{2}) \ b(k+\frac{p}{2},k-\frac{p}{2}) \ . 
\label{eq:32} 
\en 
As a warm-up exercise, we show that the pion BS-vertex is given in the 
chiral limit by 
\be 
P_0(k^2) \; = \; \frac{\sqrt{2}}{f_{\pi }} \; \gamma _5 \; \Sigma _0(k^2) 
\; , 
\label{eq:33} 
\en
where $\Sigma _0$ is the self-energy in the chiral limit, and 
$f_\pi $ is the pion decay constant. The relation (\ref{eq:33}) 
is a particular case of the general relation between the BS-equation 
at zero energy and the Schwinger-Dyson equation. From (\ref{eq:6}), one 
has that $S_0(k) \gamma_5 S_0(k) = \gamma_5 (Z_0^2(k^2) k^2 + 
\Sigma _0^2(k^2))^{-1}$. A direct calculation then yields 
\be 
K(\Sigma _0, p^2=0) \; P_0 \; = \; P_0 \; , 
\label{eq:34} 
\en 
where the Dyson-Schwinger equation (\ref{eq:5}) was used. 
$P_0$ (\ref{eq:33}) indeed describes a pseudo-scalar 
quark anti-quark bound state with zero mass, i.e $m_\pi ^2 = -p^2 =0$. 
The normalization 
of the BS-vertex is not fixed by the linear equation (\ref{eq:31}). 
It can be related to the pion-decay constant by assuming that the 
axial-vector quark anti-quark coupling is entirely dominated by the 
axial-vector pion coupling in the chiral limit~\cite{rob94}. 

We now consider the case of a small explicit breaking of the chiral 
symmetry by the current mass $m_R$. In order to calculate 
the pion mass treating the explicit chiral symmetry breaking as a 
perturbation, 
we expand the BS-equation (\ref{eq:31}) to first order\footnote{Note that 
current algebra shows that $m_\pi ^2$ is of order $m_R$, which is also 
assumed in chiral perturbation theory.} in 
$p^2 = - m_\pi ^2 $ and $m_R$, i.e. 
\be
 \left [ K(\Sigma _0, p=0) \  + \ K' (\Sigma _0, p=0) \  p^2 
\ + \delta K \right ] (P_0 \ + \ \delta P) \ = \ 
 (P_0 \ + \ \delta P) \; , 
\label{eq:35}
\en 
where $K'$ denotes the derivative of $K$ with respect to  $p^2$, and
\newline  $\delta K := K(\Sigma ,p=0) - K(\Sigma _0,p=0)$. 
The change in the pion mass in first order perturbation theory 
obtained by sandwiching equation (\ref{eq:35}) with $P_0^\dagger $. The 
contributions containing $\delta P$ drop out as a consequence of 
the BS-equation in the chiral limit, and we are left with 
\be 
0 \ =  \ - \langle P_0 \, \vert \, 
K'(\Sigma _0,0) P_0 \rangle  \ m_\pi^2  
\ + \  \langle P_0  \, \vert \, \delta K P_0 \rangle  \ . 
\label{eq:36} 
\en 
In order to evaluate the second term at the right hand side of (\ref{eq:36}), 
one uses 
\bea 
& &  ( K(\Sigma ,0) \ - \ K(\Sigma _0,0) ) P_0 = 
\nonumber \\
& & \frac{\sqrt{2}}{f_\pi } \int \frac{d^4k}{(2\pi )^4} \; 
i \gamma_\mu t^a D_{\mu \nu} (k-q) \left [
S(k) S(-k) \ - \ S_0(k) S_0(-k) \right ]  \Sigma _0(k) \gamma_5  t^a
i \gamma_\nu 
\nonumber 
\ena 
to obtain finally 
\be 
\langle P_0  \, \vert \, \delta K P_0 \rangle   \; = \; 
\frac{8N_c}{f_\pi ^2} 
\int \frac{d^4k}{(2\pi )^4} \; \Sigma _0^2(k^2)  \frac{k^2 (Z_0^2 - Z^2) + 
 \Sigma _0^2 - \Sigma ^2 }
{(Z^2 k^2 \ + \ \Sigma ^2 ) (Z_0^2 k^2 \ + \ \Sigma _0^2 )} \; . 
\label{eq:37} 
\en 
The first term at the right hand side of (\ref{eq:36}) is more involved. 
This term explicitly reads 
\bea
& & \langle P_0 \, \vert \, K'(\Sigma _0,0) P_0 \rangle \ = \
 \left . \langle P_0
 \frac{d K(\Sigma _0,p^2)}{d p^2} \right |_{p^2=0} P_0 \rangle 
\; = \; \frac{2}{ f_\pi ^2 } \; \times 
\label{eq:38} \\
& & \tr \int_{q,k} t^a i \gamma_\nu
S(q) \gamma_5 \Sigma _0(q) S(q) i \gamma_\mu t^a D_{\mu \nu}(q-k) 
\left. \frac{d}{d p^2} 
\left ( S(k_+) \Sigma _0(k) \gamma_5 S(k_-) \right )
 \right |_{p^2=0} \; , 
\nonumber 
\ena 
where $k_\pm := k \pm p/2 $. Inserting the BS-equation for the vertex 
function in the chiral limit, one gets rid of the double integration 
in (\ref{eq:38}), i.e. 
\bea 
\langle P_0 K'(\Sigma _0,0) P_0 \rangle_0 &=& \frac{2}{ f_\pi ^2 }  
\frac{d}{d p^2} \left ( \left . tr \int_k
\gamma_5  \Sigma _0(k) S_0(k_+) \gamma_5 \Sigma _0(k)
S_0(k_-)  \right ) \right |_{p^2=0} 
\label{eq:39} \\ 
&=& - 1  \; . 
\nonumber 
\ena 
The last equality was obtained with the help of equation (\ref{eq:a4}) 
of appendix \ref{app:a}. Inserting (\ref{eq:37}) and (\ref{eq:39}) into 
(\ref{eq:36}), we obtain the mass formula 
\be 
f_\pi^2 \ m_\pi^2 \ = \ 8N_c \int \frac{d^4k}{(2\pi )^4} \; 
\Sigma ^2_0(k^2)  \frac{ k^2(Z^2-Z_0^2)+\Sigma ^2 - \Sigma _0^2 }
{(Z^2 k^2 \ + \ \Sigma ^2 ) (Z_0^2 k^2 \ + \ \Sigma _0^2 )} \; . 
\label{eq:40} 
\en 
This mass formula agrees with the result found earlier in 
reference~\cite{cah96} (see also~\cite{fra96}).

\subsection{ The Gell-Mann-Oakes-Renner relation } 
\label{sec:3.2} 

Since the momentum integration at the right hand side of (\ref{eq:40}) 
is rapidly converging, one is tempted to conclude that the 
GMOR relation, i.e. 
\be 
f_\pi ^2 \, m_\pi ^2 \; = \; - 2m \, \langle \bar{q}q \rangle \; , 
\label{eq:41} 
\en 
does not hold in the context of the one-gluon-exchange models, 
because the right hand side of (\ref{eq:41}) involves a divergent 
quark condensate whereas the left hand side is finite~\cite{cah96}. 
The apparent contradiction can be resolved by taking into account 
that the bare current mass $m$ tends to zero for a large UV-cutoff 
(see (\ref{eq:21})) implying that the product of bare mass and condensate 
might be finite. 

The crucial quantity is the quark condensate which is defined via 
the trace of the quark propagator, i.e. 
\be 
\langle \bar{q}q \rangle \; = \; - \frac{N_c}{4 \pi ^2} \; 
\int _0^{\Lambda _{UV}} dk^2 \; k^2 \; \frac{ \Sigma (k^2) }{ 
Z^2(k^2) + \Sigma ^2(k^2) } \; . 
\label{eq:b1} 
\en 
In order to show that the condensate in (\ref{eq:b1}) is divergent in the 
chiral limit $m_R(\mu ) \equiv 0$, one uses the asymptotic 
behavior of the self-energy (\ref{eq:10}) and the fact that $Z(k^2)$ 
decreases to approach $1$, yielding 
\be 
\langle \bar{q}q \rangle \; = \;  \langle \bar{q} q \rangle 
(\mu ) \left[ \ln ( \mu ^2 / \laq^2 ) \right] ^{-d_m} 
 \; \left[ \ln ( \Lambda _{UV} ^2 / \laq^2 ) \right] ^{d_m} \; + \; 
\hbox{ finite terms } \; . 
\label{eq:b2} 
\en 
The crucial observation is that multiplying this divergent condensate 
with the bare current mass (see (\ref{eq:21})) results in the finite 
value 
\be 
m(\Lambda _{UV}) \, \langle \bar{q}q \rangle \; = \; 
m_R(\mu ) \, \langle \bar{q}q \rangle (\mu ) \; . 
\en 
To establish the GMOR relation (\ref{eq:41}), it remains to show that this 
finite value coincides up to a factor of $2$ with the right hand side of 
the pion mass formula (\ref{eq:40}). 
This equivalence was first observed in~\cite{la96} using the 
approximation $Z_0 \approx Z$. Below, we will generalize 
the results of~\cite{la96}. 
For this purpose, we rewrite the DS-equation for the self-energy in the 
form 
\be 
\Sigma (p^2) \; = \; m(\Lambda _{UV}) \; + \; 4 
\int _{k^2< \Lambda _{UV}} \frac{d^4k}{(2\pi )^4} \; 
\frac{ \Sigma (k^2) }{ Z^2 k^2 + \Sigma ^2 } \; d((p-k)^2) \; , 
\label{eq:b4} 
\en 
where $d(k^2)= D_{\mu \mu }(k^2) /3 $. For later convenience, we also 
introduce the DS-equation for the self-energy $\Sigma _0 $ in the chiral 
limit, i.e. 
\be 
\Sigma _0 (p^2) \; = \; 4 \int _{k^2< \Lambda _{UV}} \frac{d^4k}{(2\pi )^4} 
\; \frac{ \Sigma _0 (k^2) }{ Z_0^2 k^2 + \Sigma _0 ^2 } \; d((p-k)^2) \; , 
\label{eq:b5} 
\en 
In order to estimate the change of the self-energy due to the presence 
of the bare mass, we subtract (\ref{eq:b5}) from (\ref{eq:b4}), i.e. 
\be 
\Sigma (p^2) - \Sigma _0 (p^2) =  m(\Lambda _{UV})  +  4 
\int _{k^2< \Lambda _{UV}} \frac{d^4k}{(2\pi )^4} \; \left[ 
\frac{ \Sigma (k^2) }{ Z^2 k^2 + \Sigma ^2 } - \frac{ \Sigma_0 (k^2) }{ 
Z_0^2 k^2 + \Sigma_0 ^2 } \right] \; d((p-k)^2) \; . 
\label{eq:b6} 
\en 
We then multiply both sides of this equation with 
$$ 
\frac{ \Sigma _0 (p^2) }{ Z^2_0(p^2) p^2 \, + \, \Sigma _0(p^2) } 
$$ 
and integrate over the momentum $p$. Using the DS-equation (\ref{eq:b5}), 
the result can be cast into the form 
\bea  
&& \int _{p^2< \Lambda _{UV}} \frac{d^4p}{(2\pi )^4} \, 
\Sigma _0 (p^2) \, \frac{ \Sigma (p^2) - \Sigma _0 (p^2) }{ Z_0^2 p^2 + 
\Sigma _0^2(p^2) }  =  m(\Lambda _{UV} ) 
\int _{p^2< \Lambda _{UV}} \frac{d^4p}{(2\pi )^4} \; \frac{ 
\Sigma _0 (p^2) }{ Z_0^2 p^2 + \Sigma _0^2(p^2) } 
\nonumber \\ 
&+& 
\int _{k^2< \Lambda _{UV}} \frac{d^4k}{(2\pi )^4} \; \Sigma _0(k^2) \, 
\left[ \frac{ \Sigma (k^2) }{ Z^2 k^2 + \Sigma ^2 } - \frac{ \Sigma_0 (k^2) 
}{ Z_0^2 k^2 + \Sigma_0 ^2 } \right] \; . 
\label{eq:b7} 
\ena 
We therefore obtain the identity 
\bea 
\int _{k^2< \Lambda _{UV}} \frac{d^4k}{(2\pi )^4} && \Sigma _0(k^2) \, 
\Sigma (k^2) \left[ 
\frac{ 1 }{ Z^2 k^2 + \Sigma ^2 } - \frac{ 1 }{ 
Z_0^2 k^2 + \Sigma_0 ^2 } \right] 
\label{eq:b8} \\ 
&& = \; m(\Lambda _{UV} ) \int _{p^2< \Lambda _{UV}} 
\frac{d^4p}{(2\pi )^4} \; \frac{ 
\Sigma _0 (p^2) }{ Z_0^2 p^2 + \Sigma _0^2(p^2) } \; . 
\nonumber 
\ena 
In the product $\Sigma _0 \, \Sigma $, one might replace the 
$\Sigma $ by the self-energy $\Sigma _0 $ in the chiral limit, 
since the error is of higher order in the current mass $m_{R}$. 
Inserting then (\ref{eq:b8}) into the right hand side of the pion mass 
formula (\ref{eq:40}) reproduces the GMOR relation (\ref{eq:41}), which 
completes the proof. Recently, the GMOR relation was also shown to hold 
true in effective one-gluon exchange models models which employ a running 
bare current mass~\cite{cah96b}.

\subsection{ Numerical results } 
\label{sec:3.3} 

In the table below, we finally present numerical results for the 
pion decay constant $f_\pi $, pion mass $m_\pi $ and the quark condensate 
for five different values of the four quark coupling strength $G$. 
We also provide the results for the constituent quark mass $M_c$, 
which is defined by the pole of the constituent quark 
propagator, i.e. 
\be 
p^2 \; = \; M(p^2) \; \vert _{p^2 = M_c^2 } \; , 
\en 
where $M(p^2)$ is the mass function (\ref{eq:9a}).  

\bigskip 
\begin{tabular}{|c|c|c|c|c|} \hline 
$\frac{ G \laq ^2 }{16 \pi ^2 }$ & $M_c$ & $[- \langle \bar{q}q 
\rangle (\mu = 1 \, \hbox{GeV})]^{1/3} $ & $f_\pi $ & $m_\pi $ \\ 
\hline 
1.11 & 200 \, \hbox{MeV } & 191 \, \hbox{MeV } & 78 \, \hbox{MeV } & 
131 \, \hbox{MeV } \\ 
1.67 & 259 \, \hbox{MeV } & 203 \, \hbox{MeV } & 86 \, \hbox{MeV } & 
130 \, \hbox{MeV } \\ 
2.22 &  312 \, \hbox{MeV } & 214 \, \hbox{MeV } & 93 \, \hbox{MeV } & 
130 \, \hbox{MeV } \\ 
2.78 & 362 \, \hbox{MeV } & 223 \, \hbox{MeV } & 99 \, \hbox{MeV } & 
130 \, \hbox{MeV } \\ 
3.33 & 409 \, \hbox{MeV } & 232 \, \hbox{MeV } & 105 \, \hbox{MeV } & 
130 \, \hbox{MeV } \\ \hline 
\end{tabular} 
\bigskip 

In these calculations, $\laq \approx 500 \, $MeV was used, and 
$[m_u+m_d](\mu =1 \, \hbox{GeV}) = 15 \, $MeV for up- and down-quark masses 
was assumed. The values in the table should be compared with experimental 
ones, i.e. 
$f_\pi ^{(exp)} = 93 \, $MeV and $m_\pi ^{(exp)} = 135 \, $MeV. 
The constituent quark mass is assumed to be approximately $m_N /3 
\approx 310 \, $MeV, where $m_N$ is the nucleon mass. 
It is remarkable that the pion mass nearly stays constant when $G$ 
varies.

\section{ How effective is the NJL-model ? } 
\label{sec:4} 

\begin{figure}[t]
\centerline{ 
\epsfxsize=5cm
\epsffile{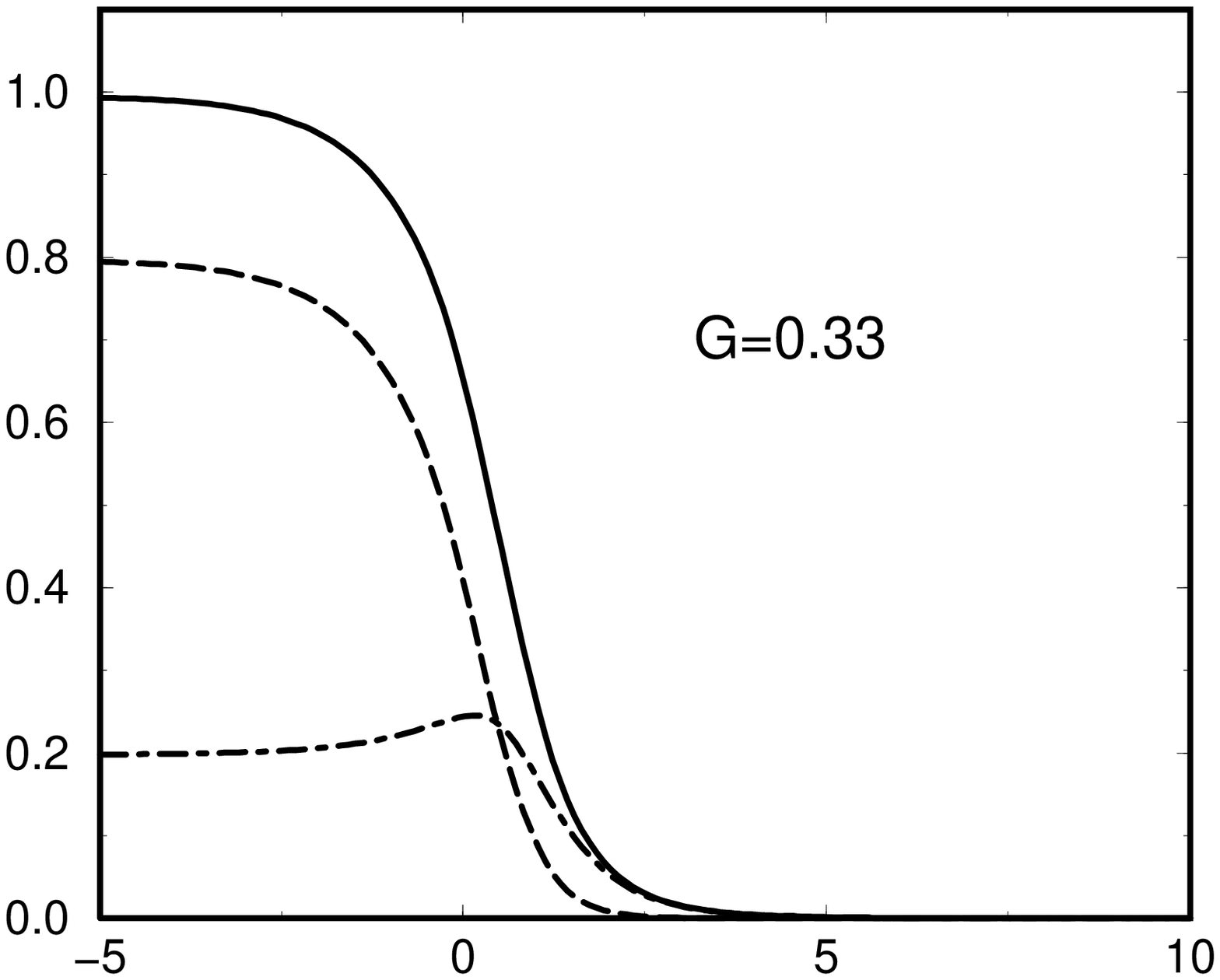} 
\epsfxsize=5cm
\epsffile{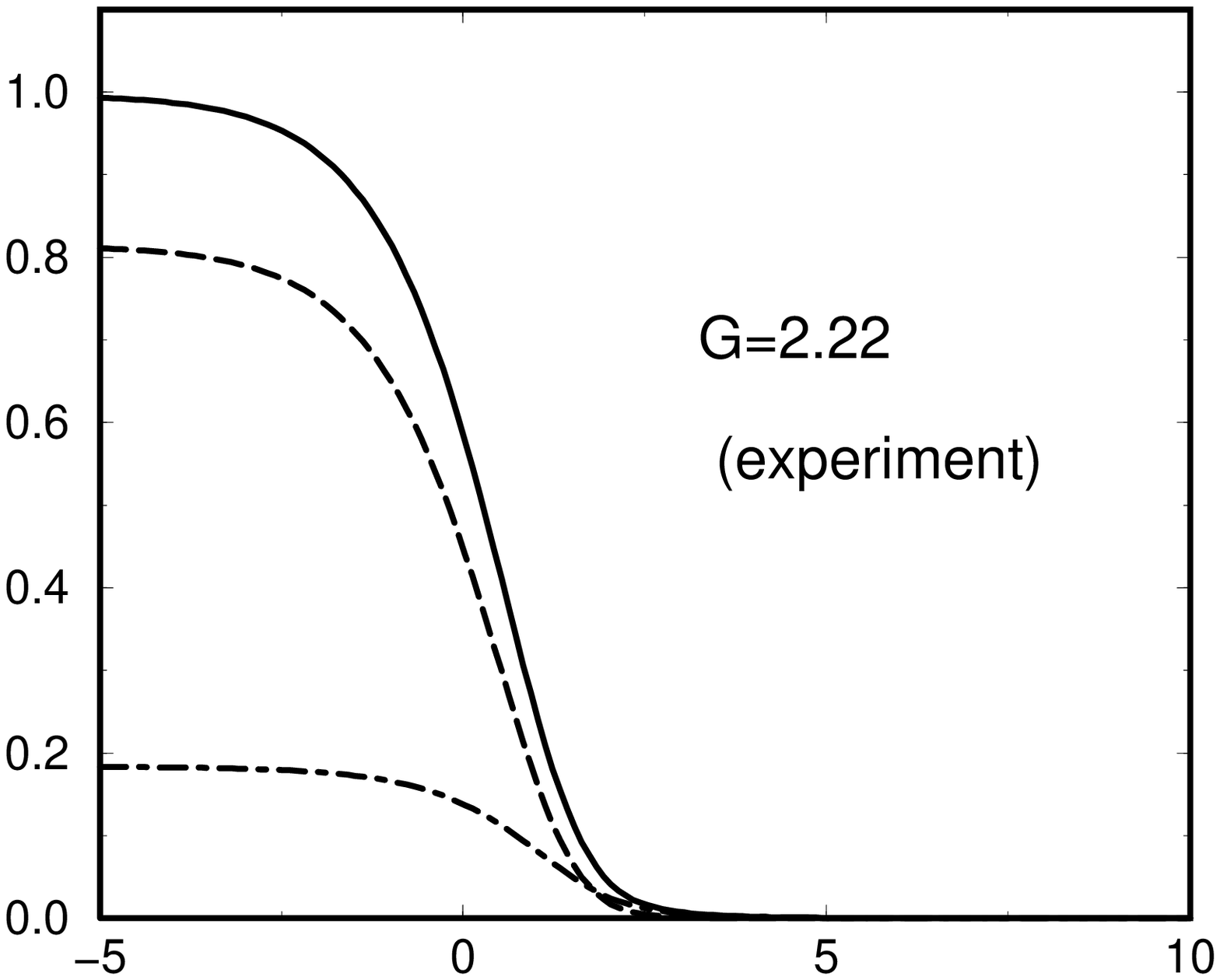} 
\epsfxsize=5cm
\epsffile{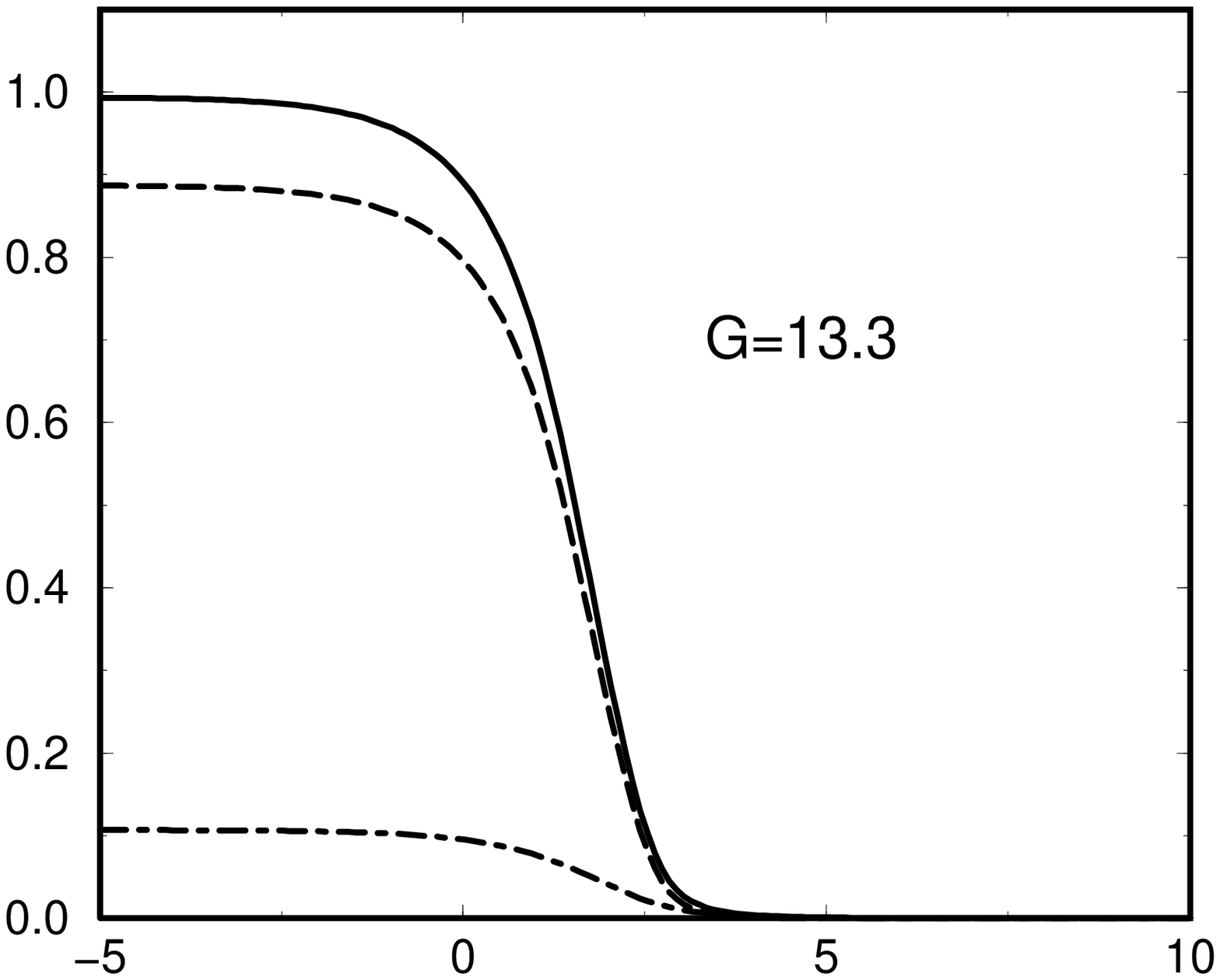} 
}
\vspace{.5cm} 
\caption{ The contribution of the NJL-term (dashed) and the contribution 
   from the non-local term (dash-dotted) to the self-energy (full) 
   as function of $\ln p^2/\laq^2 $ for various values of $G$ in units 
   of $16\pi^2 / \lan ^2$. 
   The results are rescaled to ensure $\Sigma (0)=1$. } 
\label{fig:5} 
\end{figure} 
In this section, we establish a direct relation between our 
renormalizable model and the familiar NJL-model~\cite{vo91}. 
In particular, we investigate the limits of the standard 
NJL-model~\cite{nam61} due to its non-renormalizability. 
For this purpose, we will search for a parameter range 
for which our model approximately reduces to the NJL-model. Then we 
study the accuracy of this approximation in comparison to a 
more realistic choice of parameters which reproduce the phenomenological 
data. 

The effective four quark interaction of our model consists of 
a NJL-type contact interaction at low momentum transfer $k < \lan $ 
and reproduces the effective perturbative one-gluon exchange 
at large momenta (see (\ref{eq:3})). Consequently, the self-energy 
determined by the Dyson-Schwinger equation (\ref{eq:7}) receives 
two contributions. The first term coincides (up to the momentum 
dependence of the self-energy and to the presence of $Z(k^2)$) 
with the one obtained from the NJL gap equation, while the second term 
arises from 
the one-gluon-exchange. The self-consistent solution to 
the equations for $\Sigma (k^2) $ and $Z(k^2)$ in (\ref{eq:7}) and 
(\ref{eq:8}), respectively, (see figure \ref{fig:1}) show that these 
functions are almost constant 
for $k < \lan $, and rapidly decrease at $k \approx \lan $ 
to approach the correct asymptotic behavior of $\Sigma $ as dictated 
by perturbative QCD. 
We search for a parameter range, where the second term 
on the right hand side of the DS-equation (\ref{eq:8}) only 
yields a small contribution compared with the first term. In this range, 
we can expect that observables which are insensitive to large 
loop momenta are well described by the NJL-model. The numerical analysis 
shows that such a parameter range indeed exists (see figure \ref{fig:5}). 
In the limit of a large effective four quark interaction $G$, the 
contribution to the self-energy from the non-local term is small. 
In this case, 
the approximately constant self-energy can be obtained from the reduced 
DS-equation (\ref{eq:8}), i.e. 
\be 
\Sigma (0) \; \approx \; 4G \int _{\vert k \vert < \lan } 
\frac{ d^4k }{(2\pi )^4 } \; \frac{ \Sigma (0) }{ 
Z^2(0) k^2 + \Sigma ^2(0) } \; . 
\label{eq:43} 
\en 
This equation can be cast into the gap-equation of the NJL-model, i.e. 
\be 
M \; = \; 4 G_{NJL} \; \int _{\vert k \vert < \Lambda _{NJL} } 
\frac{ d^4k }{(2\pi )^4 } \; \frac{ M }{ k^2 + M^2 } \; , 
\label{eq:44} 
\en 
where 
\be 
M \, = \, \frac{ \Sigma (0)}{Z(0)} \; , \hbo G_{NJL} \, = \, 
\frac{G}{Z^2} \; , 
\hbo 
\Lambda _{NJL} \, = \, \lan \; \approx \; 500 \, \hbox{MeV} \; . 
\label{eq:45} 
\en 
We therefore expect that observables involving rapidly converging loop 
integrals can be well estimated within the 
NJL-model, since the numerical value should be insensitive 
to the large momentum behavior. Such an observable is e.g. the 
pion decay constant defined in equation (\ref{eq:a5}) of appendix 
\ref{app:a}. 

On the other hand, the NJL-model must fail for observables 
which receive considerable contributions from large momenta of the 
loop integrals. Here, we mention the anomalous decays of the pion as an 
example. In ref.~\cite{rob94b}, it was shown that 
the rate of the decay of the neutral pion into two photons, i.e. 
\be 
\Gamma _{\pi ^0 \gamma \gamma } \; = \; 
\frac{m_\pi ^3 }{16 \pi } \, \left( \frac{ \alpha _{em} }{ 
\pi f_\pi } \right)^2 \, g \; , 
\label{eq:46} 
\en 
can be calculated in the chiral limit independently of the details of the 
quark-propagator, i.e. 
\be 
g \; = \; \int_0 ^{\Lambda ^2_{UV} \rightarrow \infty } dk^2 \; 
\frac{ \frac{d}{dk^2} C}{ (1+C(k^2))^3 } \; = \; 
\int _0 ^{\infty } dC \frac{1}{(1+C)^3} \; = \; \frac{1}{2} \; , 
\label{eq:47} 
\en 
where $C(k^2):= \Sigma ^2 (k^2) / (k^2 Z^2(k^2) ).$ The agreement of the 
theoretical prediction with the experimental value $g^{exp} = 0.504 $ is 
very good and supports the constituent quark picture of hadrons.  
Since the formula (\ref{eq:47}) is independent of the details 
of the model, it is straightforward to apply it to the case of the 
NJL-model. In this case, the momentum integration in (\ref{eq:47}) is 
cutoff at $\Lambda _{NJL}$, and $C(k^2) = M^2/k^2 $. One  
obtains\footnote{ This result holds for an O(4) invariant sharp cutoff. 
Different regularization prescriptions might lead to different values $g$.}
\be 
g \; = \; \int_0 ^{\Lambda ^2_{NJL} } dk^2 \; \frac{ \frac{d}{dk^2} C}{ 
(1+C(k^2))^3 } \; = \; \int _{M^2/\Lambda _{NJL}^2} ^{\infty } 
dC \frac{1}{(1+C)^3} \; = \; \frac{1}{2} \, \frac{ 1 }{ 
1 + \frac{ M^2 }{ \Lambda _{NJL }^2 } } \; . 
\label{eq:48} 
\en 
The typical value of the ratio of constituent quark mass and NJL-cutoff 
is of order $M/\Lambda _{NJL } \approx 1/2$ implying that 
in the NJL-model the $g$-factor is underestimated by 20\%. 

The phenomenological data (see the table in the previous section) 
require an interaction strength $G \approx 2.22 \times 16 \pi^2 / \lan ^2$. 
For this value, the NJL-term contributes 82\% to the full 
self-energy. This result suggests that hadronic observables calculated 
within a NJL-model, where $\Lambda _{NJL} $ is tightly related to the 
interaction strength $G_{NJL}$ by (\ref{eq:4a}), should deviate from the 
corresponding 
results obtained by a renormalizable model by approximately 18\%. 
This is indeed the typical accuracy of the predictions of the 
NJL-model~\cite{vo91}. 

Finally, we mention that the NJL-propagator which arises as a limit 
of our renormalizable model possesses the structure 
\be 
S_{NJL} (k) \; = \; \frac{i}{ Z \, \kslash \; + \; i \Sigma  } \; , 
\hbox to 2cm {\hfil with \hfil } Z \approx 2.3 
\label{eq:49} 
\en 
(despite of the generic form of the gap equation (\ref{eq:44})). 
In the standard NJL-model, the $Z$-factor is independent of the momentum 
and set to $1$ in order to obtain the generic 
residuum of the quark propagator. 
On the other hand in renormalizable models, the $Z$-factor depends, 
in principle, on the interaction strength. Fortunately, our numerical 
calculations have shown (see figure \ref{fig:3}) that $Z$ 
is approximately constant 
for sufficiently large coupling strength $G$, which is the parameter 
region where our model can be approximately reduced to an NJL-model. 
Therefore our investigations suggest a NJL-model with a momentum cutoff 
(see equation (\ref{eq:45})), .i.e. 
$\Lambda _{NJL} = \lan \approx 500 \, $MeV. This value is in fact 
slightly smaller than the typical 
value, which arises from the fit of pion data within the NJL-model 
when pseudo-scalar axial-vector meson mixing is disregarded~\cite{vo91}.

\section{Summary and conclusions } 

We have proposed an effective quark theory which can be regarded 
as an (one loop) renormalizable extension of the phenomenologically 
successful NJL-model. The effective quark interaction of 
this model is given by a NJL-contact interaction at low momentum transfer 
$(k \leq \lan )$, and is mediated by the one-gluon exchange 
augmented by the logarithmic QCD corrections at high energies, i.e. 
$$ 
D _{\mu \nu } (k) \; = \; 
\left[ G \, \theta (\lan - \vert k \vert ) \; + \; 
\frac{4 \pi \alpha (k^2) }{ k^2 } \theta (\vert k \vert - \lan ) \right] 
\, \left( \delta _{\mu \nu } - \frac{ k_\mu k_\nu }{k^2} \right) \; , 
$$ 
where 
$$ 
G \; = \; \frac{4 \pi \alpha (\lan ^2) }{ \lan ^2 } \; , \; 
\alpha (k^2) \; = \; \frac{12 \pi }{ (33-2 N_f) \ln k^2/\laq ^2} \; . 
$$ 
The intermediate energy scale $\lan $ at which the low energy interaction 
matches the high energy one gluon exchange, is fixed by the choice of $G$. 
We demand $\lan > \laq $ in order to account for the screening of the 
Landau pole~\cite{la95b}. The model is (at least) one loop renormalizable. 
The numerical and analytical study of the asymptotic behavior of the quark 
self-energy $\Sigma (k^2)$ has 
yielded the correct anomalous dimensions for the current mass and 
quark condensate, which are know from perturbative QCD calculations. 
A chiral phase transition from the perturbative chirally symmetric 
phase to a non-trivial chirally broken phase characterized by a 
non-vanishing quark condensate was found at sufficiently strong 
interactions, i.e. for 
$$
G \; > \; G_c \approx \biggl( 0.26 \pm 0.011 \biggr) 
\; \frac{16 \pi ^2 }{ \laq ^2 } \; . 
$$
The numerical data for the quark mass function and the quark condensate 
for an interaction strength $G$ close to the critical one, $G_c$, are well 
fitted by 
\bea 
M (p^2=0) & \approx & 0.43 \, \laq \; \left( \frac{G}{G_c}-1 
\right)^{0.47} \; , 
\nonumber \\ 
- \langle \bar{q} q \rangle (\mu = 1\, \hbox{GeV}) & \approx & 
0.041 \; \laq ^3 \; \left( \frac{G}{G_c} -1 \right) ^{0.27} \; . 
\nonumber 
\ena 
The critical exponent for the self-energy is consistent with 
$1/2$, whereas the exponent of the quark condensate significantly 
deviates form $1/2$. 

Within our model, the emergence of the pion as a Goldstone boson was 
studied. The Bethe-Salpeter equation for the pion vertex 
was solved using a derivative expansion. We have 
analytically verified the Gell-Mann-Oakes-Renner relation (\ref{eq:41}), 
which relates the product of pion mass and decay constant to the product 
of current mass and condensate. 

We have also established the relation of our model with the 
familiar NJL-model thereby exhibiting the limits of the latter due 
to its non-renormalizability. Like the NJL-model, our model does 
not contain quark confinement. We believe, however, that quark 
confinement is not of much relevance for low-energy hadron 
dynamics (if the external momenta are much smaller than the 
quark thresholds). On the other hand, renormalizability 
seems to be much more important, in particular, for processes which 
are sensitive to large loop momenta as e.g. the anomalous decay of the 
neutral pion. Our model is obviously superior to the 
NJL-model in these cases.

\appendix 
\section{ Numerics } 
\label{app:a0} 

\parindent=0pt
{\bf 1. Solving the Dyson-Schwinger equation } 

In order to solve the coupled set of Dyson-Schwinger equations 
(\ref{eq:7},\ref{eq:8}), the integration over the angle 
(between the vectors $p$ and $k$) and over the loop momentum $k$ 
must be done numerically. 
For this purpose, we have discretized the momentum at $n$ points in 
logarithmic steps and used a cubic spline interpolation for the self-energy 
$\Sigma (k^2)$ and the quark wave-function $Z(k^2)$. 
The coupled set of differential equations is then solved by iteration. 
We used $n=20$ and $n=30$. It turned out that $n=20$ is sufficient for 
most of the applications. For numerical reasons, the momentum integration 
is cut off at $\Lambda _{UV}^2 = 3.3 \times 10^6 \laq^2$. 
By using also $\Lambda _{UV}^2 = 6.6 \times 10^7 \laq^2$, we have checked 
that our results are practically independent of the numerical cutoff. 
Since the functions $\Sigma $ and $Z$ depend on $(p+k)^2$, the 
numerical program needs an evaluation of this functions up to twice 
the numerical cutoff. Since the functions are only available for 
momenta smaller than $\Lambda ^{UV}$, we used 
\be 
\{Z, \, \Sigma \} (k^2) \; = \; \{Z, \, \Sigma \} (\Lambda _{UV}^2)  
\hbox to 2 true cm{\hfil for \hfil } k^2 > \Lambda _{UV}^2 \; . 
\label{eq:a01} 
\en 
This procedure is part of the regularization scheme. In the scaling limit, 
the physical results are independent of the details described in 
(\ref{eq:a01}). However, it turns out that the definition limits 
the numerical accuracy for the values of $\Sigma (k^2)$, $Z(k^2)$ for 
large values of $k^2$. In order to estimate statistical and 
systematic numerical errors, we set up a second program, where 
we solved the Dyson-Schwinger equations (\ref{eq:7},\ref{eq:8}) {\it after} 
the loop momentum $k \rightarrow k-p$ was shifted. In the second program, 
one need not to resort to a truncation scheme like that in (\ref{eq:a01}). 

From subsection \ref{sec:2.3}, we know that the self-energy $\Sigma (k^2)$ 
decreases like $1/k^2 $ up to logarithmic corrections. Close to the 
numerical cutoff, the self-energy takes values which are $10^{-7}$ times 
smaller than the self-energy at zero momentum. In order to 
extract the slope of the logarithmic decrease (which is related to the 
quark condensate by (\ref{eq:10})), a refinement of the numerical 
approach is necessary. It turned out to be convenient to introduce 
the auxiliary variable 
\be 
\sigma (k^2) \; = \; \left( 1 \, + \, \frac{ k^2 }{ \laq ^2 } \right) 
\; \Sigma (k^2) \; . 
\label{eq:a02} 
\en 
At small values of $k^2$, $\sigma (k^2)$ essentially coincides with 
$\Sigma (k^2) $. The advantage is that $\sigma (k^2)$ decreases only 
logarithmically at asymptotic large values of $k^2$. 
To summarize, the second program solves the equivalent set of 
differential equations, i.e. 
\bea 
\sigma (p^2) &=& m \, + \, 4G \int _{\vert k -p \vert < \lan } 
\frac{ d^4k }{(2\pi )^4 } \; \frac{ \sigma (k^2) }{ 
N(k^2) } 
\frac{ \laq ^2 + p^2 }{ \laq ^2 + k^2 } 
\label{eq:a03} \\ 
&+& 4 \int _{\vert k-p \vert > \lan } 
\frac{ d^4k }{(2\pi )^4 } \; \frac{ \sigma (k^2) }{ N(k^2) } 
\frac{ \laq ^2 + p^2 }{ \laq ^2 + k^2 } \; 
\frac{ 4 \pi \alpha ((k-p)^2) }{ (k-p)^2 } \; , 
\nonumber \\ 
Z(p^2) &=& 1 + \frac{4G}{3} \int _{\vert k-p \vert < \lan } 
\frac{ d^4k }{(2\pi )^4 } \, \frac{ Z (k^2) }{ N(k^2) } 
\left( 3 \frac{p \cdot k }{p^2} + 2 \frac{ (p \cdot k)^2 - 
p^2 k^2 }{ (k-p)^2 } \right) 
\nonumber \\ 
&+& \frac{4}{3} \int _{\vert k -p \vert > \lan } 
\frac{ d^4k }{(2\pi )^4 } \; \frac{ Z (k^2) }{ N(k^2) } 
\left( 3 \frac{p \cdot k }{p^2} + 2 \frac{ (p \cdot k)^2 - 
p^2 k^2 }{ (k-p)^2 } \right) \, 
\nonumber \\ 
&\times & \frac{ 4 \pi \alpha ((k-p)^2) }{ (k-p)^2 } \; , 
\label{eq:a04} 
\ena 
where 
\be 
N(k^2) \; = \; Z^2(k^2) k^2 + \frac{ \sigma ^2(k^2) }{1 + k^2/\laq^2 } 
\; . 
\en 
It turns out that both programs produce practically the same results, 
which establishes that our results are independent from the 
regularization scheme. The second program is, however, superior 
for calculating the quark condensate. 

\medskip 
{\bf 2. The quark condensate } 

The quark condensate can be extracted from the asymptotic behavior 
of the quark mass function in (\ref{eq:10}). For this purpose, 
we exploit the fact that the function $M(p^2) p^2 $ is almost a 
linear function of $\ln ( \ln p^2)$ for sufficiently large values 
of $p^2$ (see figure 2). A line fit provides the quark condensate and 
the numerical estimate of the anomalous dimension $D_m$. In order to 
estimate the error of the condensate, we use equation (\ref{eq:b2}), 
which analytically relates the trace of the quark propagator to the 
quark condensate. We learn that the error $d_m-D_m$ in the determination 
of the anomalous dimension implies that we miss the true value for 
the condensate by a factor 
\be 
\left[ \ln \frac{ \Lambda _{UV}^2 }{ \laq ^2 } \right] ^{d_m-D_m} \; . 
\label{eq:a05} 
\en 
Since we know the correct anomalous dimension analytically 
($d_m= 4/9$ from (\ref{eq:11}) and (\ref{eq:21})), equation 
(\ref{eq:a05}) allows for an estimate of the numerical error, and 
gives rise to the error bars in figure \ref{fig:4}. 

\medskip 
{\bf 3. The critical coupling } 

In order to extract the critical coupling $G_c$ (see equation 
(\ref{eq:25})), we consider the self-energy (see figure \ref{fig:3})) 
and the quark condensate (see figure \ref{fig:4}), respectively, 
at small values $G-G_c$. We perform a mean square fit to the $6$--$8$ 
points which are closest to $G_c$. For this range of the coupling constant, we 
expect that the quantities $\Sigma $ and $\langle \bar{q} q \rangle $ 
show the critical behavior according 
\be 
\{ \Sigma , \langle \bar{q} q \rangle \} \; = \; 
c_{\Sigma, \langle \bar{q} q \rangle } \, \left( \frac{G}{G_c} -1 
\right) ^{\delta _{\Sigma , \langle \bar{q} q \rangle } } \; . 
\label{eq:a06} 
\en 
We then perform a mean square fit according (\ref{eq:a06}) to the self-energy 
and the condensate data points, respectively. This procedure, in particular, 
results in the critical exponents $\delta _\Sigma $ and $\delta 
_{\langle \bar{q} q \rangle }$ and in two (nearly identical) values 
for $G_c$. The difference in the obtained values for $G_c$ in each case 
provides an estimate of the numerical error in $G_c$.

\section{ Asymptotic form of the self-energy } 
\label{app:aa} 

In order to estimate the contribution of the gluon exchange term 
(second line of (\ref{eq:7})), we shift the momentum integration, 
yielding 
\bea 
\Sigma (p^2) &\approx & m \; + \; 
4 \int \frac{ d^4k }{(2\pi )^4 } \; \frac{ \Sigma (k) }{ 
Z^2 k^2 + \Sigma ^2 } \; \frac{ 4 \pi \alpha ((p-k)^2) }{ (p-k)^2 } 
\theta (\vert p-k \vert - \lan ) \; , 
\label{eq:14} \\ 
Z(p^2) & \approx & 1 \; + \; \frac{4}{3 p^2} 
\int \frac{ d^4k }{(2\pi )^4 } \; \frac{ Z(k) }{ 
Z^2 k^2 + \Sigma ^2 } \; \left(  2 p \cdot k + \tr \, 
\frac{ \pslash (\kslash - \pslash ) \kslash (\kslash - \pslash ) }{ 
4 (p-k)^2 } \right) 
\label{eq:14a} \\ 
&& \frac{ 4 \pi \alpha ((p-k)^2) }{ (p-k)^2 } 
\theta (\vert p-k \vert - \lan ) \; . 
\nonumber 
\ena 
We also employ a modified Landau approximation, 
i.e. 
\be 
\alpha \left( (p-k)^2 \right) \, \theta \left( (p-k)^2 - \lan ^2 \right) \; 
\approx \; 
\alpha \left( \hbox{max}(p^2,k^2) \right) \, \theta \left( \hbox{max}(p^2,k^2)
- \lan ^2 \right) \; . 
\label{eq:15} 
\en 
Numerical results confirm that this approximation is valid in the 
context under considerations of this section. The technical benefit 
is that we can now explicitly 
perform the angle integration in (\ref{eq:14}) and (\ref{eq:14a}). 
In particular, the angle-integral in (\ref{eq:14a}) corresponds 
to that of QED in Landau gauge, which is known to be zero~\cite{co89}. 
It is therefore consistent to set $Z(p^2) =1$ for the large values of 
$p^2$, in agreement with the numerical results (see figure \ref{fig:1}). 
Within the approximation (\ref{eq:15}) the asymptotic momentum 
dependence of the mass function $M(p^2)$ coincides with that of 
the self-energy $\Sigma (p^2)$, i.e. 
\be 
M(p^2) \; \approx \; \Sigma (p^2) \; , \hbox to 2cm {\hfil for \hfil } 
p^2 \gg \Sigma ^2 (0) \; . 
\label{eq:16} 
\en 
It is straightforward to evaluate the angle integration in (\ref{eq:14}), 
i.e. 
\be 
\Sigma (p^2) \approx m \; + \; \frac{ \alpha (p^2) }{\pi p^2 } 
\int _0^{p^2} dk^2 \; k^2 \; \frac{ \Sigma (k^2) }{ k^2 + \Sigma ^2 } 
\; + \; \frac{1}{\pi } \int _{p^2}^{\Lambda _{UV}} dk^2 \; 
\alpha (k^2) \, \frac{ \Sigma (k^2) }{ k^2 + \Sigma ^2 } \; . 
\label{eq:17} 
\en 
where an UV-cutoff $\Lambda _{UV} $ was introduced, that will be 
taken to infinity at the end. 

We will now show that the Ans\"atze (\ref{eq:12}) and (\ref{eq:12a}) 
are self-consistent 
solutions to (\ref{eq:17}) at large momentum $p^2$. 
First, the case of explicit chiral symmetry breaking is considered. 
To calculate the first integral at the right hand side of 
(\ref{eq:17}), an arbitrary but sufficiently large scale $\nu$ 
was introduced so that the self-energy $\Sigma (p^2)$ can be 
approximated by (\ref{eq:12})) for $p^2 \gg \nu ^2$, i.e. 
\bea 
\frac{ \alpha (p^2) }{\pi p^2 } && 
\int _0^{p^2} dk^2 \; k^2 \; \frac{ \Sigma (k^2) }{ k^2 + \Sigma ^2 } 
\; = \; \frac{ \alpha (p^2) }{\pi p^2 } \int _0^{\nu^2} dk^2 \; 
k^2 \; \frac{ \Sigma (k^2) }{ k^2 + \Sigma ^2 } 
\label{eq:18} \\ 
&+& \frac{ \alpha (p^2) }{\pi p^2 } \int _{\nu^2}^{p^2} dk^2 \; 
k^2 \; \frac{ \Sigma (k^2) }{ k^2 + \Sigma ^2 } 
\; = \; \frac{ 12 \kappa }{ (33 - 2N_f) \, \ln ^{\alpha +1} p^2/\laq ^2 } 
\nonumber \\ 
&+& {\cal O} \left( \frac{ \kappa }{ \ln ^{\alpha +2} p^2/\laq ^2 } 
\right) \; + \; {\cal O } \left( \frac{ \nu ^2 }{ p^2 
\ln p^2/\laq ^2 } \right) \; . 
\nonumber 
\ena 
The second integral on the right hand side of (\ref{eq:17}) is 
\bea  
\frac{ 12 }{ 33 -2 N_f} && \int _{p^2}^{\Lambda _{UV}^2} \frac{du}{u} 
\frac{ \kappa }{ \ln ^{\alpha +1} u/ \laq ^2 } \; = \; 
\label{eq:19} \\ 
&& \frac{ 12 }{ (33 -2 N_f) \; \alpha } \left[ 
\frac{ \kappa }{ \ln p^2 / \laq^2 } - \frac{ \kappa }{ 
\ln \Lambda _{UV} ^2 / \laq^2 } \right] \; . 
\nonumber 
\ena 
We find that the term (\ref{eq:18}) is sub-leading at large $p^2$. 
Our Ansatz solves (\ref{eq:17}) for 
\be 
\alpha \; = \; \frac{12}{33 - 2N_f} \; = \; d_m \; , \hbo \kappa \; = \; 
m_R (\mu ) \left[ \ln ( \mu ^2 / \laq^2 ) \right] ^{d_m} \; , 
\label{eq:20a} 
\en 
and a bare current mass 
\be 
m(\Lambda _{UV}) \; = \; \frac{ m_R (\mu ) \left[ \ln ( \mu ^2 / \laq^2 ) 
\right] ^{d_m} }{ \ln ^\alpha \Lambda _{UV}^2 
/ \laq ^2 } \; . 
\label{eq:21a} 
\en 
Our goal is now to show that $\alpha $ can be identified with the 
anomalous dimension of the current mass, and that then 
the high-energy behavior of the 
quark mass function as predicted by the operator product expansion 
(see (\ref{eq:10})) is obtained. 

In the absence of explicit chiral symmetry, we assume equation 
(\ref{eq:12a}) to be the correct high momentum behavior of the 
self-energy $\Sigma (p^2)$. One first estimates 
\bea 
&& \frac{ \alpha (p^2) }{\pi p^2 } \, 
\int _0^{p^2} dk^2 \; k^2 \; \frac{ \Sigma (k^2) }{ k^2 + \Sigma ^2 } 
\; = \; \frac{ \alpha (p^2) }{\pi p^2 } \int _0^{\nu^2} dk^2 \; 
k^2 \; \frac{ \Sigma (k^2) }{ k^2 + \Sigma ^2 } 
\label{eq:22} \\ 
&& + \frac{ \alpha (p^2) }{\pi p^2 } \int _{\nu^2}^{p^2} dk^2 \; 
k^2 \; \frac{ \Sigma (k^2) }{ k^2 + \Sigma ^2 } 
\; = \; d_m \frac{ c }{ (1-\beta ) \; p^2 \, \ln ^{\beta } p^2/\laq ^2 } 
\nonumber \\ 
&& + {\cal O } \left( \frac{ 1 }{ p^2 \ln p^2/\laq ^2 } \right) \; . 
\nonumber 
\ena 
Here $\beta $ had to be assumed smaller than $1$, which will later turn 
out to be the case. 

The last term on the right hand side of (\ref{eq:17}) is 
\be 
\frac{1}{\pi } \int _{p^2}^{\Lambda _{UV}^2} dk^2 \; \alpha (k^2) \, 
\frac{ \Sigma (k^2) }{ k^2 + \Sigma ^2 } 
\; = \; {\cal O } \left( \frac{ 1 }{ p^2 \ln ^{\beta +1 } 
p^2/\laq ^2 } \right) \; . 
\label{eq:23} 
\en 
In contrast to the case of explicit chiral symmetry breaking (\ref{eq:19}), 
the contribution of the second term (\ref{eq:23}) is now 
sub-leading to the term (\ref{eq:22}) (provided that $0 < \beta <1 $). 
The contribution from the upper bound $\Lambda _{UV}^2 $ in 
(\ref{eq:23}) decreases faster than $1/ \Lambda _{UV}^2$. We can 
therefore safely remove the cutoff, i.e. $\Lambda _{UV}^2 \rightarrow 
\infty $, since this contribution decreases faster with the regulator 
than counter-terms. 
This, in particular, implies that we must have $m_R \equiv 0$, since 
the contribution from the regulator can not account for the logarithmic 
decrease with $\Lambda _{UV}$ of the current mass term. In order the 
Ansatz (\ref{eq:12a}) to be a solution of the 
Dyson-Schwinger equation in the asymptotic region (\ref{eq:17}), 
one must impose 
\be 
\beta \; = \; 1 - d_m \; , \hbo m_R(\laq ) \equiv 0 \; . 
\label{eq:24} 
\en 
Note that for three quark flavors $N_f=3$ one finds $\beta = 0.555 \ldots $ 
and $0 < \beta <1 $ as assumed before. 
Our main observation here is that the Ansatz (\ref{eq:12a}) is the 
correct asymptotic behavior of the self-energy 
in the chiral limit, accompanied by the correct 
anomalous dimensions of the quark condensate as obtained by the 
operator product expansion (see (\ref{eq:10})).

\section{ Pion decay constant and normalization of the BS-vertex} 
\label{app:a} 

The normalization of the pion BS-vertex in (\ref{eq:33}) can be
derived from the observation that the axial-vector vertex $\Gamma_5^\mu$ 
contains a pion pole term~\cite{ja73}. With the assumption that
this pole term dominates in the chiral limit, 
we have~\cite{rob94} 
\be 
\vec{\Gamma}_5^\mu(k_+,k_-)  \stackrel {\longrightarrow}{ {}_{p \rightarrow 0}}
\sum _a P^a (k_+,k_-) ( \frac{1}{p^2}) \ \langle 0 | \vec{A}_5^\mu 
| \pi ^a(p) \rangle
\  \stackrel {\longrightarrow}{ {}_{p \rightarrow 0}} \vec{P}(k_+,k_-) 
f_\pi \ \frac{p^\mu}{p^2} \; , 
\label{eq:aa1} 
\en 
where the arrows indicate the iso-spin vector. 
Combing this with the Ward identity for the axial vector vertex
in the chiral limit
\be
p_\mu \vec{\Gamma}_5^\mu (k_+,k_-) \ = \ 
\frac{\vec{\tau}}{2} \ \left [ S^{-1}(k_+) \gamma_5 + \gamma_5 S^{-1}(k_-)
 \right ] 
\label{eq:aa2} 
\en 
one recovers that the quark self-energy constitutes the BS-vertex in the 
chiral limit. In addition, (\ref{eq:aa2}) provides its proper 
normalization, i.e. 
\be
\vec{P_0}(k,0) \ = \ \vec{\tau} \ \gamma_5 \frac{\Sigma_0(k)} {f_\pi} \ .
\label{eq:aa3} 
\en 
A particular pion state, e.g. the $\pi^+$ which we will consider in the
following, is given by 
\be
P^+_0(k) \ = \frac{1}{\sqrt{2}} \left ( P^1_0(k) + i P^2_0(k) \right )
 = \  \tau^+ \ \gamma_5 \frac{\sqrt{2}}{f_\pi} \Sigma_0(k) \ , \quad 
 \tau^+ = \frac{1}{2} (  \tau_1 + i \tau_2 ) \ . 
\label{eq:aa4} 
\en 

In order to extract an explicit formula for the pion decay constant 
$f_\pi $, we exploit the fact that the pion electro-magnetic charge 
is properly normalized to one. For this purpose, we study the 
electro-magnetic form-factor at zero photon momentum, i.e. 
\bea 
2 p_\mu F_{\pi}(p^2,0) &=& 
\tr \left [  \frac{2}{3} \int_k P_0^\dagger(k) S(k_+)
\Gamma_\mu(k_+,k_+) S(k_+) P_0(k) S(k_-) \right . 
\label{eq:a1} \\ 
&-&  \frac{1}{3} \int_k 
\left . P_0^\dagger(k) S(k_+) 
  P_0(k) S(k_-) \Gamma_\mu(k_-,k_-) S(k_-) \right ] \; , 
\ena 
where $k_\pm = k \pm p/2$, and $\Gamma_\mu(k_-,k_-) $ is the full 
quark photon vertex. This vertex is unambiguously provided by the 
differential form of the Ward identity, i.e. 
\be 
 \Gamma_\mu(k,k) \; = \; \frac{\partial}{\partial k^\mu} S^{-1}(k) \; . 
\label{eq:a2} 
\en 
With the help of this equation, (\ref{eq:a1}) can be cast into 
\be 
2 \,  p_\mu \; = \; - 2 \, \tr \left [ \frac{2}{3} \int_k 
 P_0^\dagger(k)  \frac{\partial S(k_+)}{\partial p^\mu} P_0(k) S(k_-)  
\ +  \frac{1}{3} \int_k  P_0^\dagger(k)
 S(k_+) P_0(k) \frac{\partial S(k_-)}{\partial p^\mu} \right ]
  \  . 
\label{eq:a3} 
\en 
Using the decomposition $S(p) =  i \pslash \sigma _V (p^2) + \sigma _S 
(p^2) $ for the quark-propagator (\ref{eq:6}), one realizes that 
both integrals at the right hand side of (\ref{eq:a3}) yield 
the same value. One therefore finds 
\be 
1 \; = \; - \, \frac{d}{d p^2} \left ( \left . tr \int_k
P_0^\dagger(k) S_0(k_+)  P_0(k) S_0(k_-)  
\right ) \right |_{p^2=0} \; . 
\label{eq:a4} 
\en 
Inserting the BS-vertex (\ref{eq:33}) in the chiral limit into 
(\ref{eq:a4}), we finally obtain an explicit expression for the 
pion decay constant, i.e. 
\bea 
f_\pi ^2 & = & \frac{N_c}{8 \pi^2} \int 
 \Sigma _0^2 \left [ - 2 \sigma_S \sigma'_S + \sigma^2_V 
- 2 \sigma_V \sigma_V' k^2 - k^2 \sigma_S \sigma''_S 
\right. \label{eq:a5} \\ 
&+& \left. k^2 \sigma_S' \sigma'_S - k^4 \sigma_V \sigma_V'' 
+ k^4 \sigma_V' \sigma_V' \right ] k^2 d k^2 \; , 
\nonumber 
\ena 
where the prime denotes the derivative of the function with respect 
to its argument $k^2$. This formula was first obtained in~\cite{ben95}.

\bigskip 
{\bf Acknowledgments: } 

We thank R.\ Alkofer for valuable comments on the manuscript.

\end{document}